\documentclass[a4paper,10pt]{article}
\usepackage{url}
\usepackage[a4paper,margin=3.5cm]{geometry}
\usepackage[english]{babel}
\usepackage{graphicx}
\usepackage{amsmath}
\usepackage{color}
\usepackage[usenames,dvipsnames]{xcolor}
\usepackage{amssymb}
\usepackage{morefloats}
\usepackage{footnote}
\usepackage{multicol}
\usepackage{hyperref}
\usepackage{enumerate}
\usepackage[symbol]{footmisc}
\usepackage{rotfloat}
\restylefloat{figure}
\makeatletter
\DeclareOldFontCommand{\rm}{\normalfont\rmfamily}{\mathrm}
\DeclareOldFontCommand{\sf}{\normalfont\sffamily}{\mathsf}
\DeclareOldFontCommand{\tt}{\normalfont\ttfamily}{\mathtt}
\DeclareOldFontCommand{\bf}{\normalfont\bfseries}{\mathbf}
\DeclareOldFontCommand{\it}{\normalfont\itshape}{\mathit}
\DeclareOldFontCommand{\sl}{\normalfont\slshape}{\@nomath\sl}
\DeclareOldFontCommand{\sc}{\normalfont\scshape}{\@nomath\sc}
\makeatother
\begin{document}

\title{Comparison of unfolding methods using RooFitUnfold}
\author{Lydia Brenner\footnote{Corresponding author.}, Pim Verschuuren$^*$,  \\Rahul Balasubramanian, Carsten Burgard, Vincent Croft,  \\Glen Cowan, Wouter Verkerke}


\date{\today}
\maketitle
\thispagestyle{empty}


\begin{abstract}

In this paper we describe RooFitUnfold, an extension of the RooFit statistical software package to treat unfolding problems, and which includes most of the unfolding methods that commonly used in particle physics. The package provides
a common interface to these algorithms as well as common uniform methods to evaluate their performance in terms of bias, variance and coverage. In this paper we exploit this common interface of RooFitUnfold to compare
the performance of unfolding with the Richardson-Lucy, Iterative Dynamically Stabilized, Tikhonov, Gaussian Process, Bin-by-bin and inversion methods on several example problems.

\end{abstract}



\section{Introduction}
\label{sec:intro}

In High Energy Physics (HEP) and in many other fields one often
measures distributions of quantities such as particle energies or
other characteristics of observed events.  Because the experimental
apparatus (the ``detector'') inevitably has a limited resolution, the
measured (or ``reconstructed'') value of the quantity in question will
differ in general from its true value.  This results in a distortion
or smearing of the measured distribution relative to what would be
obtained if the detector had perfect resolution.  The statistical
procedure of estimating the true distribution from the directly
measured one is usually called unfolding in HEP, or deconvolution in
many other fields.  Unfolding algorithms and their implementation in
software have been widely discussed in HEP (see, e.g.,
Refs.~\cite{Cowan98,zech2016analysis,Cowan:2002in,kuusela2015,refId0,Blobel:2011fih,Behnke:2013pga}).

In this paper we describe an extension of the statistical software
RooFit \cite{theRooFitpackage} to treat unfolding problems called
RooFitUnfold.  The unfolding algorithms implemented in the package and
studied here include: the Richardson-Lucy algorithm
\cite{Richardson,Lucy,dagostini}, Iterative Dynamically Stabilized
\cite{bogdan1} and Tikhonov \cite{Tikhonov77,hocker,tackmann}
unfolding, as well as unregularised and bin-by-bin unfolding.
RooFitUnfold provides a common interface to the algorithms and to
methods for evaluating their performance.

In Sec.~\ref{sec:formulation} we provide a mathematical description of
the unfolding problem and discuss various concepts that enter into the
algorithms that are studied.  Section~\ref{sec:framework} then
presents the specific algorithms implemented in RooFitUnfold and also
describes how a number of important quantities of interest (bias,
variance, coverage probability) are found using the software.  In
Sec.~\ref{sec:results} we illustrate use of the different algorithms
by means of several examples that compare properties of the unfolded
distributions.  Conclusions are given in Sec.~\ref{sec:discussion}.


\clearpage
\section{Mathematics of the unfolding problem}
\label{sec:formulation}

A brief description is provided here of the mathematics behind the
unfolding algorithms studied in this paper.  A definition of the
unfolding problem is presented in Sec.~\ref{sec:definition} and some
methods for its solution are described in
Sec.~\ref{sec:unfoldingmethods}.  These methods require one to choose
the degree of smoothness imposed on the solution, as discussed in
Sec.~\ref{sec:reg}.




\subsection{Definition of the unfolding problem}
\label{sec:definition}

The unfolding problem is formulated here largely following the
notation and terminology of Ref.~\cite{Cowan98} (Ch.~11).  Consider a
counting experiment where a histogram of some variable $x$ is used to
construct a histogram.  Let $\vec{\mu} = (\mu_1, \ldots, \mu_M)$ be
the expected number of events in $M$ bins that would be found if the
variable $x$ could be measured exactly for each event.  We will call
this the ``true histogram'', and the goal of unfolding is to estimate
these $M$ parameters.  We will use hats to denote the estimators,
i.e., $\hat{\vec{\mu}} = (\hat{\mu}_1, \ldots, \hat{\mu}_M)$.

Alternatively one may wish to estimate the probabilities

\begin{equation}
\label{eq:probi}
p_i = \frac{\mu_i}{\mu_{\rm tot}} \;,
\end{equation}

\noindent where $\mu_{\rm tot} = \sum_{i=1}^M \mu_i$ is the total
expected number of events.  Below we will focus on the case where the
goal is to estimate the $M$ components of the true histogram
$\vec{\mu}$.

Suppose that for each event the measured value of the variable $x$
differs in general from the true value by some random amount owing to
the limited resolution of the detector.  Let us suppose that the
measured values are used to construct a histogram $\vec{n} = (n_1,
\ldots, n_N)$, whose bins may in general differ in their boundaries
and total number $N$ relative to the $M$ bins of the true histogram.
These data values follow some probability distribution, often taken to
be Poisson or approximately Gaussian, and have expectation values
$\vec{\nu} = E[\vec{n}] = (\nu_1, \ldots, \nu_N)$.  We will refer to
$\vec{n}$ as the data histogram (or simply ``the data'') and
$\vec{\nu}$ as the ``expected data histogram''.

The expected data histogram $\vec{\nu}$ has a form that differs from
what would be obtained with perfect resolution since events with a
true value of the variable $x$ in a given bin may be measured in a
different one.  This migration of events may be expressed as (see,
e.g., Ref.~\cite{Cowan98})

\begin{equation}
\label{eq:nuRmu}
\nu_i = \sum_{j=1}^M R_{ij} \mu_j \;,
\end{equation}

\noindent where $i = 1, \ldots, N$ and the response matrix,

\begin{equation}
\label{eq:responsemat}
R_{ij} = P(\mbox{measured value in bin } i|\mbox{true value in bin }j) \;,
\end{equation}

\noindent gives the conditional probability for an event to be
reconstructed in bin $i$ of the data histogram given that its true was
in bin $j$ of the true histogram.  The response matrix can be
determined using simulated events combined with Monte Carlo modelling
of the detector's response. It is to first approximation dependent
only on the properties of the measurement device, but it can depend at
some level on the physics model used to generate the simulated events.
For purposes of the studies presented in this paper we will suppose
that the response matrix is known with negligible error, and we focus
here on the uncertainties that result as a consequence of the
unfolding procedure itself.

The effect of migration between bins is to smear out peaks or fine
structure in $\vec{\mu}$ resulting in a flatter histogram for
$\vec{\nu}$.  The goal of unfolding is to estimate the true histogram,
i.e., the parameters $\vec{\mu} = (\mu_1, \ldots, \mu_M)$ using the
data values $\vec{n} = (n_1, \ldots, n_N)$ combined with the knowledge
of the detector's response through the matrix $R$.  The data
are described with a given probability model that determines
the likelihood function $L(\vec{\mu}) = P(\vec{n} | \vec{\mu})$.
Often the $n_i$ are modelled as independent and Poisson distributed,
so that the likelihood is

\begin{equation}
  \label{eq:poissonlnl}
  L(\vec{\mu}) = \prod_{i=1}^N \frac{\nu_i^{n_i}}{n_i!} e^{- \nu_i} \;,
\end{equation}

\noindent where $\nu_i = \sum_{j=1}^M R_{ij} \mu_j$.  Alternatively,
the observed number of events may be modelled as following a Gaussian
distribution with mean $\nu_i$ and standard deviation $\sigma_{n_i}$.
In this case the log-likelihood is (up to an additive constant) a sum
of squares,

\begin{equation}
\label{eq:gausslnl}
  \ln L(\vec{\mu}) = - \frac{1}{2} \sum_{i=1}^N \frac{(n_i -
    \nu_i)^2}{\sigma_{n_i}^2} \;.
\end{equation}

\noindent If the $N \times M$ response matrix $R$ is in fact square, i.e., equal
numbers of bins for both $\vec{\mu}$ and $\vec{\nu}$, and if it is
nonsingular, then $\vec{\mu} = R^{-1} \vec{\nu}$.  In this case one
can take the estimators for $\vec{\mu}$ to be

\begin{equation}
\label{eq:matrixinv}
\hat{\vec{\mu}} = R^{-1} \vec{n} \;.
\end{equation}

\noindent This follows (see, e.g., Ref.~\cite{Cowan98}) from the
likelihoods above for which the maximum-likelihood estimators for the
$\nu_i$ are $\hat{\nu}_i = n_i$.

The most important properties of the estimators that we investigate
here are the bias (difference between estimator's expectation and
parameter's true values),

\begin{equation}
\label{eq:bias}
b_i = E[\hat{\mu}_i] - \mu_i \;,
\end{equation}

\noindent and the covariance matrix $\mbox{cov}[\hat{\mu}_i,
  \hat{\mu}_j]$, in particular its diagonal elements (the variances)
$V[\hat{\mu}_i]$ (also written $\sigma_{\hat{\mu}_i}^2$).  In the
  Poisson case with a square nonsingular response matrix, the
  maximum-likelihood estimators have zero bias and their covariance
  equals the minimum variance bound (see, e.g., Ref.~\cite{Cowan98}).
  Nevertheless the variances are extremely large and in this sense the
  unfolding problem is said to be ill-posed
  \cite{Hansen:1990:DPC:98694.98703}.  In unfolding one therefore
  attempts to reduce the variance through some form of regularisation,
  and this necessarily introduces some bias.

\subsection{Solutions to the unfolding problem}
\label{sec:unfoldingmethods}

To suppress the large variance of the maximum-likelihood estimator one
chooses a $\hat{\vec{\mu}}$ that does not correspond to the maximum of
the log-likelihood $\ln L_{\rm max}$, but rather one considers a
region of $\vec{\mu}$-space where $\ln L(\vec{\mu})$ is within some
threshold below $\ln L_{\rm max}$, and out of these the distribution
is chosen that is smoothest by some measure.  This can be achieved by
maximising not $\ln L(\vec{\mu})$ but rather a linear combination of
it and a regularisation function $S(\vec{\mu})$, which represents the
smoothness of the histogram $\vec{\mu}$.  That is, the estimators
$\hat{\vec{\mu}}$ are determined by the maximum of

\begin{equation}
  \label{eq:regular}
  \varphi(\vec{\mu}) = \ln L(\vec{\mu}) +  \tau S(\vec{\mu}) \;,
\end{equation}

\noindent where the regularisation parameter $\tau$ determines the
balance between the two terms.  Equivalently one can take the
regularisation parameter to multiply $\ln L(\vec{\mu})$, as
done, e.g., in Ref.~\cite{Cowan98}.

The regularisation function $S(\vec{\mu})$ can be chosen in a number
of ways (see, e.g., Refs.~\cite{Cowan98,zech2016analysis}).  Two of
the algorithms studied here (TUnfold and SVD; see
Sec.~\ref{sec:overview}) use Tikhonov regularisation \cite{Tikhonov77}
based on the mean squared second derivative of the unfolded
distribution.  For discrete bins $\vec{\mu}$, the Tikhonov
regularisation function can be expressed as (see, e.g.,
\cite{Cowan98})

\begin{equation}
\label{finite_diff2}
S(\vec{\mu}) = - \, \sum_{i=1}^{M-2} (-\mu_{i} \, 
+ \, 2 \mu_{i+1} \, - \,
\mu_{i+2})^{2} .
\end{equation}

\noindent If the regularisation parameter $\tau$ is set to zero one obtains the
maximum-likelihood estimators (the ``unregularised solution'',
cf.\ Sec.~\ref{sec:roounfoldinvert}), which have zero (or small) bias
but very large statistical variance.  In the limit of large $\tau$ the
resulting $\hat{\vec{\mu}}$ is the maximally smooth distribution
corresponding to the maximum of $S(\vec{\mu})$.  This is independent
of the data and thus has zero statistical variance, but results in a
large bias.

Other algorithms to construct regularised estimators do not involve
maximisation of a function such as the $\varphi(\vec{\mu})$ in
Eq.~(\ref{eq:regular}).  For example, the Richardson-Lucy
\cite{Lucy,Richardson} iterative method, e.g., as implemented by
D'Agostini \cite{dagostini} (see Sec.~\ref{sec:iterativebayes}), a
trial solution is successively updated.  The number of updates plays
the role of the regularisation parameter, such that zero iterations
gives the maximally smooth trial solution and for a large number the
estimators tend towards those of maximum likelihood.  In the method
based on Gaussian Processes (GP) by Bozson et al.\ \cite{bozson} (see
Sec.~\ref{sec:gp}) the regularisation is set through the kernel
function of the GP.  For essentially all unfolding methods the analyst
must choose, explicitly or otherwise, some parameter that regulates
the degree of smoothness imposed on the solution and thus determines
the trade-off between bias and variance in the estimators
$\hat{\vec{\mu}}$.

\subsection{Determining the regularisation parameter}
\label{sec:reg}

Regularised unfolding thus requires a choice of algorithm by, for
example, selecting a regularisation function $S(\vec{\mu})$, as well
as some prescription for setting the degree of regularisation, e.g.,
through the parameter $\tau$ for Eq.~(\ref{eq:regular}) or the number
of iterations used in the Lucy-Richardson algorithm.  The
regularisation parameter determines the trade-off between bias and
statistical variance in the estimators $\hat{\vec{\mu}}$ that result.
A quantity that represents this trade-off is the mean squared error
(sum of bias squared and variance) averaged over the bins,

\begin{equation}
  \label{eq:msedef}
  \mbox{MSE} = \frac{1}{M} \sum_{i=1}^M \left( V[\hat{\mu}_i]  +  b_i^2
  \right) \;.
\end{equation}

\noindent Minimising the (bin-averaged) MSE as defined above is one of
the criteria investigated here for setting the regularisation
parameter.

Another criterion that can be employed to set the degree of
regularisation is based on the coverage probability of confidence
intervals constructed for each bin of the true histogram, as proposed
in Ref.~\cite{kuusela2015}.  These can be taken as extending between
plus and minus one standard deviation about the estimator, i.e.,
$[\hat{\mu}_i - \sigma_{\hat{\mu}_i}, \hat{\mu}_i +
  \sigma_{\hat{\mu}_i}]$.  Such an interval will contain the true
value $\mu_i$ with a specified coverage probability and we take the
average of these,

\begin{equation}
  \label{eq:coverage}
  P_{\rm cov} = \frac{1}{M} \sum_{i=1}^M
  P( \hat{\mu}_i - \sigma_{\hat{\mu}_i} < \mu_i <
\hat{\mu}_i + \sigma_{\hat{\mu}_i} | \vec{\mu}) \;,
\end{equation}

\noindent as a criterion that can be used to choose the regularisation
parameter.  For the unregularised estimators ($\tau \rightarrow 0$),
$P_{\rm cov}$ tends towards the nominal coverage for a confidence
interval of plus-or-minus one standard deviation about the
maximum-likelihood estimator (assuming it is Gaussian distributed),
i.e., $P_{\rm nom} = 68.3\%$.  As the amount of regularisation is
increased, the standard deviations $\sigma_{\hat{\mu}_i}$ decrease,
and thus the coverage probability also decreases.  The recipe used in
this paper is to select the regularisation parameter ($\tau$ or the
number of iterations) such that $P_{\rm cov} = P_{\rm nom} -
\varepsilon$ with the threshold $\varepsilon = 0.01$.

\clearpage
\section{The RooFitUnfold framework}
\label{sec:framework}

The RooFitUnfold framework provides a unified software environment in which various unfolding algorithms used in HEP are implemented. The framework currently implements all of the algorithms available in the RooUnfold package~\cite{phystat2011}, in addition to the Gaussian Process unfolding method. The unique proposition of the RooFitUnfold framework is that it generalises the description of input distributions for unfolding problems as RooFit~\cite{theRooFitpackage} probability models, whereas RooUnfold and most standalone unfolding implementation only accept histograms as input distribution.

The use of binned probability models as input extends the functionality of unfolding procedures to problems where uncertainties other than statistical uncertainties are important and must be propagated to the unfolded distribution.  Such uncertainties can be generically expressed in terms of nuisance parameters that affect the binned physics distribution that is intended to be unfolded. Common uncertainties in HEP that are captured in nuisance parameters include uncertainties from simulation statistics and systematic uncertainties originating from theory and detector modelling. In this paper we will not make use of RooFitUnfold's ability to propagate nuisance parameters, but will exploit its unified implementation for the evaluation of the bias and variance of unfolding methods, allowing a comparison of the unfolding algorithms on equal footing. 

Presently, RooFitUnfold has no specific technical handling of empty bins and low-statistics bins at the framework level, but will warn the user of (potentially) problematic input ingredients if the response matrix is not invertible, or
if this inversion results in numerical stability problems. The choice of binning is considered to be responsibility of the user.

Section~\ref{sec:overview} will provide a brief overview of the unfolding methods implemented, Section~\ref{sec:usingmc} explains the role of MC simulation in determining unfolding ingredients, and Section~\ref{sec:bvest} details the exact procedure that are used for the estimation of bias and variance of these
methods in RooFitUnfold, which will be used in the results presented in Section~\ref{sec:results}/.

\subsection{Overview of unfolding methods}
\label{sec:overview}

In this overview a brief description of each method is given, with references to the general unfolding methodology presented in Sec.~\ref{sec:formulation} where appropriate.
Further details on each algorithm can be found in the references cited.

\subsubsection{Richardson-Lucy (Iterative Bayes)}
\label{sec:iterativebayes}
The RooUnfoldBayes algorithm uses the iterative method described by
D'Agostini in \cite{dagostini}.  Starting from equal probabilities in
all bins, the solution is updated using a rule based on Bayes'
theorem.  In other fields this algorithm is known as Richardson-Lucy
deconvolution \cite{Richardson,Lucy} and is typical of truncated
expectation maximisation algorithms
\cite{Dempster77maximumlikelihood}.
The regularisation strength corresponds to the number of iterations
performed, with the solution approaching the maximum-likelihood
estimator as the number of iterations increases.

\subsubsection{Singular Value Decomposition (SVD)}
\label{sec:svd}
The routine RooUnfoldSvd provides an interface to the TSVDUnfold class
implemented in ROOT by Tackmann \cite{tackmann}, which uses the
Tikhonov unfolding method \cite{Tikhonov77} in the manner described by
H\"ocker and Kartvelishvili \cite{hocker}.  The data are modelled as
Gaussian distributed resulting in a log-likelihood given by a sum of
squares as in Eq.~(\ref{eq:gausslnl}).  Singular Value Decomposition
is used to express the detector response as a linear series of
coefficients.  The regularisation parameter, equivalent to $\tau$ in
Eq.~(\ref{eq:regular}), can be related to the singular values of the
response matrix.


\subsubsection{TUnfold}
\label{sec:tunfold}
RooUnfoldTUnfold provides an interface to the TUnfold method
implemented by Schmitt \cite{schmitt}.  TUnfold uses Tikhonov
regularisation and the log-likelihood is taken to have the
sum-of-squares form of Eq.~(\ref{eq:gausslnl}).
TUnfold can automatically determine an optimal regularisation
parameter by scanning the `L-curve' \cite{Lawson1974}.

\subsubsection{Iterative Dynamically Stabilized (IDS)}
\label{sec:ids}

The iterative, dynamically stabilized (IDS) unfolding method \cite{bogdan1}, is an iterative unfolding method that has
a regularisation method based on the statistical significance of the difference
between observed and simulated data. RooUnfoldIds uses a reweighting procedure that aims
to stabilise the unfolding procedure against feature of the data that
are not present in the model.

\subsubsection{Gaussian Process (GP)}
\label{sec:gp}
The unfolding method of Bozson et al.\ \cite{bozson}, as implemented in RooUnfoldGP, constructs the
estimator for the true histogram as the mode of a posterior
probability obtained using Bayesian regression.  For
Gaussian-distributed data the estimator for the true histogram is
equivalent to the mean function of a Gaussian Process (GP) conditioned
on the maximum likelihood estimator.  The kernel function of the GP
introduces regularisation, which has a natural interpretation as the
covariance of the underlying distribution. This approach allows for
the regularisation
to be varied along the distribution.


\subsubsection{Bin-by-bin}
\label{sec:binbybin}
The bin-by-bin method implemented in RooUnfoldBinByBin defines
estimators of the form $\hat{\mu}_i = C_i n_i$, where the `correction
factors' are computed for each bin using a Monte Carlo simulation of
the true events and the detector's response as $C_i = \mu_i^{\rm MC} /
\nu_i^{\rm MC}$.  Further details and drawbacks of this method are
discussed in Ref.~\cite{Cowan98}.

\subsubsection{Matrix Inversion}
\label{sec:roounfoldinvert}
For the case of equal bins in the true and measured histograms ($M =
N$) and assuming a nonsingular response matrix $R$ one can
construct the unregularised estimators of Eq.~(\ref{eq:matrixinv}).
RooUnfoldInvert performs inversion of the response matrix with
singular value decomposition (TDecompSVD).  The estimators have larger
variances than from any regularised method but zero conceptual 
bias.  This solution can be useful, e.g., if there is relatively
little migration of events between bins so that the response matrix is
almost diagonal.  Even if in the case of greater migration and thus
larger statistical variances, the unregularised solution can be used
together with the full covariance matrix of the estimators to carry
out a meaningful statistical test of a hypothetical true distribution.


\subsection{Determining unfolding ingredients with Monte Carlo simulation}
\label{sec:usingmc}

Simulated data from Monte Carlo (MC) models are used in RooFitUnfold
to determine the response matrix $R$ defined in
Eq.~(\ref{eq:responsemat}), which is needed for all of the unfolding
methods except the bin-by-bin method of Sec.~\ref{sec:binbybin}.  In
addition, MC data are used to obtain model predictions for the true
histogram $\vec{\mu}$ and the expected data histogram $\vec{\nu}$.

In a full Monte Carlo simulation in HEP one generates events each
characterised by a true value of the variable in question $x_{\rm
  true}$ according to a given physical theory such as the Standard
Model.  The response of the detector is then also simulated resulting
in a measured value $x_{\rm meas}$.  Suitably normalised histograms of
the $x_{\rm true}$ and $x_{\rm meas}$ values are used to determine the
model's predictions for the true and expected data histograms
$\vec{\mu}$ and $\vec{\nu}$.  It is often the case that the MC samples
are sufficiently large that we may neglect statistical uncertainties
in these quantities.

The MC simulated events are also used in RooFitUnfold to determine the response
matrix $R$.  In practice the simulated events are used first to find
the transfer matrix $N_{ij}$ defined as the number of events found
with $x_{\rm meas}$ in bin $i$ and $x_{\rm true}$ in bin $j$.  The
response matrix is then

\begin{equation}
  \label{eq:responsemat2}
  R_{ij} = \frac{N_{ij}}{\sum_{k=1}^N N_{kj}} \;.
\end{equation}

\noindent A sufficiently large sample of simulated data and
appropriate choice of binning is important to ensure that the response
matrix determined in this way is sufficiently smooth and not overly
influenced by statistical fluctuations.

\subsection{Bias, variance and coverage estimation procedures}
\label{sec:bvest}
The calculation of the bias and variance of the estimators is based on
a given choice for the true distribution $\vec{\mu}$, and in general
these quantities can depend on this choice.  For a given $\vec{\mu}$
one computes the expected observed histogram $\vec{\nu} = R
\vec{\mu}$, and a simulated data sample $\vec{n}$ is generated with
the $n_i$ independent and Poisson distributed having mean values
$\nu_i$.  We refer to this as a ``toy'' MC data set, $\vec{n}_k$,
where $k = 1,\ldots,K$.  In the studies shown below we have used $K =
1000$.  Each dataset $\vec{n}_k$ is unfolded with the chosen method to
acquire the estimates $\hat{\vec{\mu}}_k$. The variance for bin $i$ is
then calculated as

\begin{equation}
  \sigma^2_{\hat{\mu_{i}}} = \frac{1}{K-1}\sum_{\rm k}^K
  \left( (\hat{\mu_{i}})_k - \frac{1}{K}\sum_{\rm k}(\hat{\mu}_i)_k \right)	
\end{equation}

\noindent Using the same datasets the bias for bin $i$ is calculated
as


\begin{equation}
	b_{i} = \frac{1}{K}\sum_{\rm k}^K(\hat{\mu_{i}})_k - \mu_{i} \;.
\end{equation}

Under the assumption that the $\hat{\mu}_i$ are Gaussian distributed,
the coverage probability $P_\textrm{cov}$ of the intervals
$[\hat{\mu}_{i} - \sigma_{\hat{\mu}_{i}}, \hat{\mu}_{i} +
  \sigma_{\hat{\mu}_{i}}]$, defined in Eq~(\ref{eq:coverage}), can be
calculated in closed form~\cite{kuuselaphd} using $\sigma_{\mu_{i}}$
and $b_{i}$ as

 \begin{equation}
\label{eq:covest}
P_{\textrm{cov}} =
\Phi\Big(\frac{b_{i}}{ \sigma_{\hat{\mu}_{i}}} + 1\Big) -
  \Phi\Big(\frac{b_{i}}{\sigma_{\hat{\mu}_{i}}} - 1\Big),
\end{equation}

 \noindent where $\Phi$ is the Standard Gaussian cumulative
 distribution function.

\section{Comparison of methods}
\label{sec:results}


In this section we present a comparison of the performance of the unfolding methods described in Section~\ref{sec:overview}. 
While realistic physics models and detector response functions in HEP rely on elaborate MC simulation packages that simulate both the physics and
detector response from first principles, we opt in this study for simple and analytically expressed physics functions and resolution models. These analytical models represent a 
more portable benchmark than the elaborate simulation models, while still introducing a realistic level of complexity. Sections~\ref{sec:expsm} and Sections~\ref{sec:expbsm} present
results on an example model that is representative of unfolding problems commonly encountered in HEP, and is evaluated in situations where the response matrix
sampled from the same or a different model, respectively, as the expected data. Section~\ref{sec:bimodal} will further explore the onset of bias for scenarios in which
model of the expected data and the response matrix disagree, using a distortable bimodal distribution that is smeared with a Gaussian resolution model. 

\subsection{Study on smeared exponential distribution - SM} 
\label{sec:expsm}

\begin{figure}[!ht]
\noindent \begin{centering}
\includegraphics[width=.549\textwidth]{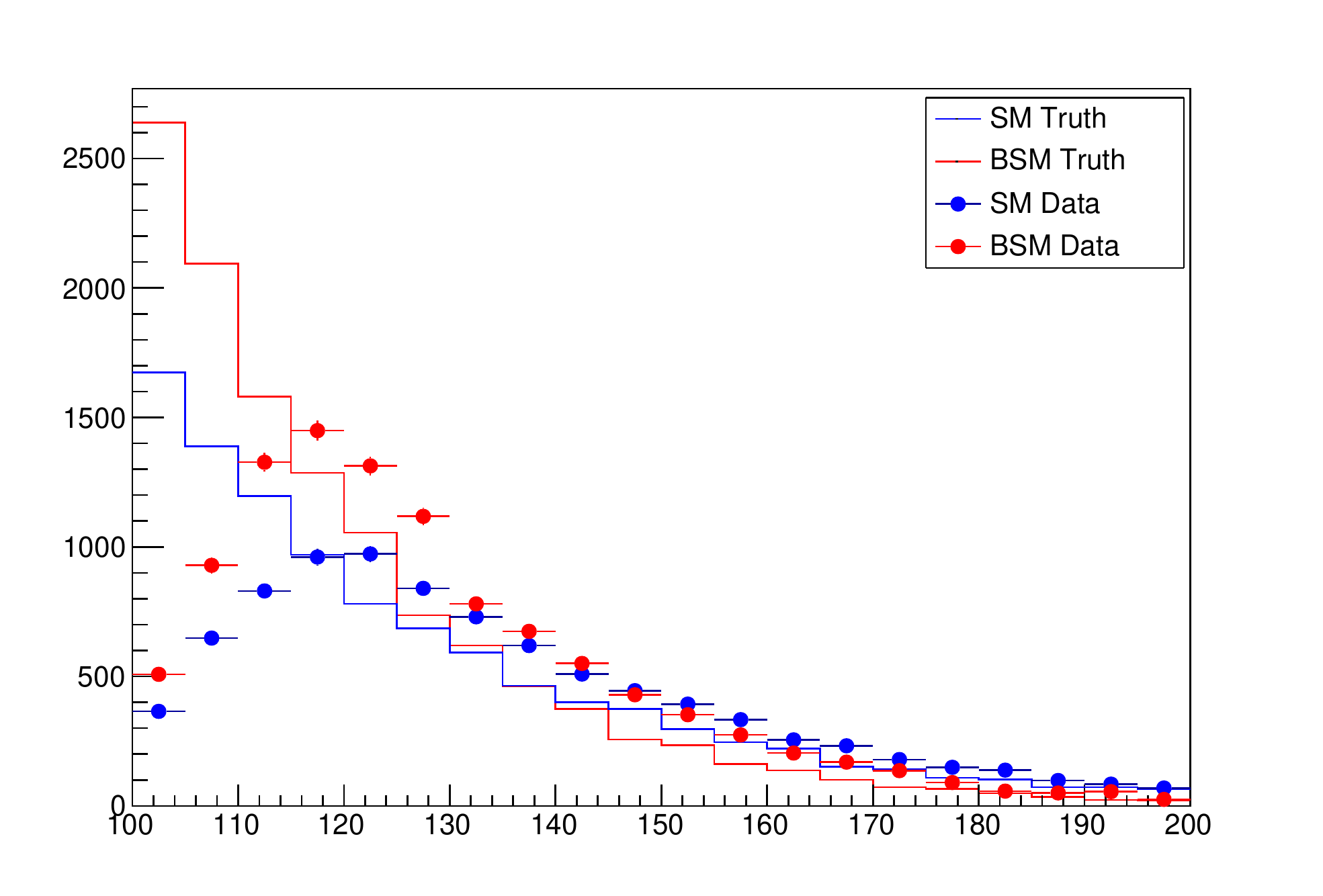}
\includegraphics[width=.431\textwidth]{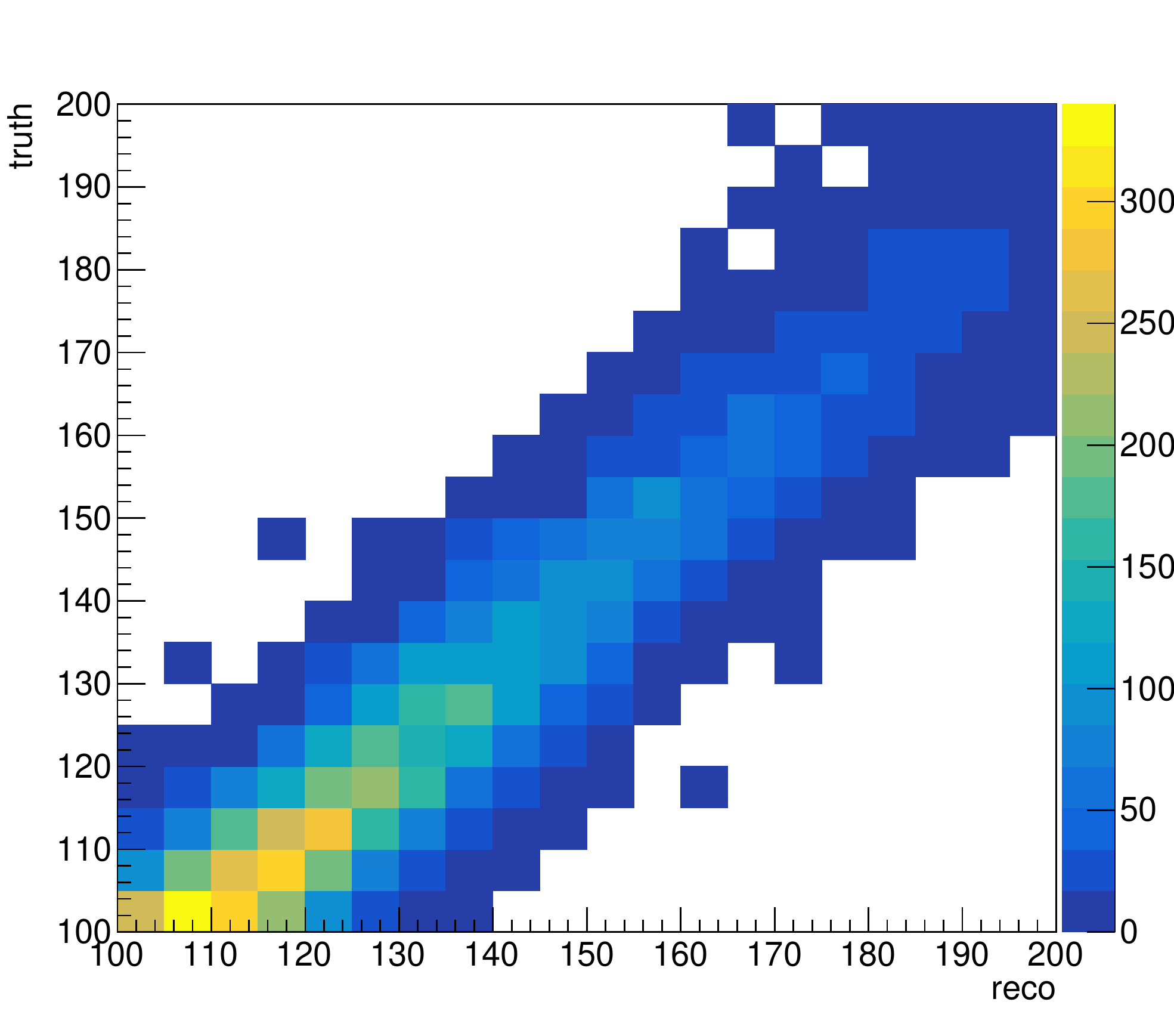}
\par\end{centering}
\caption{Left: The true distributions for the SM and BSM models and the corresponding smeared datasets. Right: the transfer matrix for the SM model, which is populated by the same events as the SM true distribution shown.}
\label{fig:model_exp}
\end{figure}

We define as benchmark model an exponential decay distribution, smeared with a resolution function that is loosely inspired on a calorimeter response:

\begin{eqnarray*}
f(x|\alpha) & = & f_\textrm{physics}(x_\textrm{true}|\alpha) \ast f_\textrm{detector}(x_\textrm{true},x) \\
            & = & \left( \alpha \cdot \exp(-\alpha  \cdot x_\textrm{true}) \right) \ast \textrm{Gauss}\left(x-x_\textrm{true},7.5,  0.5 \cdot \sqrt{x_\mathrm{true}}+2.5\right),
\end{eqnarray*}

\noindent where the $\ast$ symbol represents the convolution operator. We define two models variants, labeled SM (`Standard Model') and BSM (`Beyond the Standard Model'), that correspond to an exponential distribution with a slope $\alpha$ of 0.035 and 0.05 respectively. The true and expected distributions $\vec{\mu}$ and $\vec{\nu}$ corresponding to this model are defined by 20 uniformly sized bins in the range $[100,200]$.
The true distributions and data for SM and BSM, populated with 10000 and 14000 events respectively, are shown in Fig.~\ref{fig:model_exp}, along with the SM transfer matrix.

We first perform a standard unfolding benchmark using only the SM model: we compare for all methods the unfolded distributions of the SM data, obtained using the response matrix populated with data sampled from the same SM model, to the SM true distribution. 
In the top row of Fig.~\ref{fig:bigplot_sm} the unfolded data is shown for every algorithm, along with the SM true distribution. Each panel also separately visualises the variance and bias of the unfolded distribution for each bin, calculated with the procedure described in Section~\ref{sec:bvest} as well as corresponding coverage calculated with Eq.~\ref{eq:covest}. In the visualisation of the bias for each bin, the statistical error $\sigma_{\hat{\mu}}$ is also shown to help guide the interpretation of observed fluctuations in the
unfolded data. 
For each algorithm with a tuneable regularisation strength (Iterative Bayes, IDS, TSVD, TUnfold) two unfolding solutions are shown. These correspond to a regularisation strength that
\begin{enumerate}
\item unconditionally minimises the bin-averaged MSE , or
\item additionally requires the that the bin-averaged estimate of the coverage of the unfolded data reaches the target coverage of 68.3\% within 1\%.
\end{enumerate}
The bottom row of Fig.~\ref{fig:bigplot_sm} shows the average MSE and average coverage, i.e. the optimisation criteria used, as function of the regularisation strength parameter\footnote{NB: Although all plots are shown in the same format, the practical realisation and range of the regularisation strength parameter is different for each algorithm.} 
Vertical lines indicate the chosen regularisation strengths that correspond to two solutions shown in the top pane. For the Gaussian Process and Bin-by-Bin algorithms, which have no tuneable regularisation parameter, 
as well as for the unregularised Matrix Inversion method only a single solution is shown. 

We observe that all tuneable methods (Iterative Bayes, IDS, TVSD, TUnfold) can reproduce the target distribution well, with an average bias consistent with zero in all bins. However, the solutions
corresponding to the unconditionally minimised bin-averaged MSE substantially undercover for all four algorithms. Imposing the condition of correct bin-averaged coverage in the tuning largely rectifies
this situation and results in a nearly uniformly correct bin-by-bin coverage, with only handful of bins undercovering by a few percent. 

The Gaussian Process algorithm, which internally tunes its regularisation strength, performs well in regions far away from the lower observable boundary ($x>120$) with uniformly good coverage and no significant bias.
In the regime close to the boundary there is significant bias and undercoverage, which may be caused by boundary leakage effects related to the default choice of kernel function and might be mitigated with a different choice of kernel function. 
The unregularised Matrix Inversion method exhibits perfect coverage, but has a variance that is many orders of magnitude larger than that of any of the other algorithms, whereas the performance
of the Bin-by-Bin method in terms of bias, variance and significant undercoverage is comparable to that of the tuneable methods.

\subsection{Study on smeared exponential distribution - BSM} 
\label{sec:expbsm}

\begin{sidewaysfigure*}[p]
\begin{tabular}{cccccc}
 & & & & & Matrix Inversion \& \\
 Bayes & 
 IDS & 
 TSVD & 
 TUnfold & 
 Gaussian Process &
Bin-by-Bin  \\
\includegraphics[trim=14 0 55 0,clip,width=.155\textwidth]{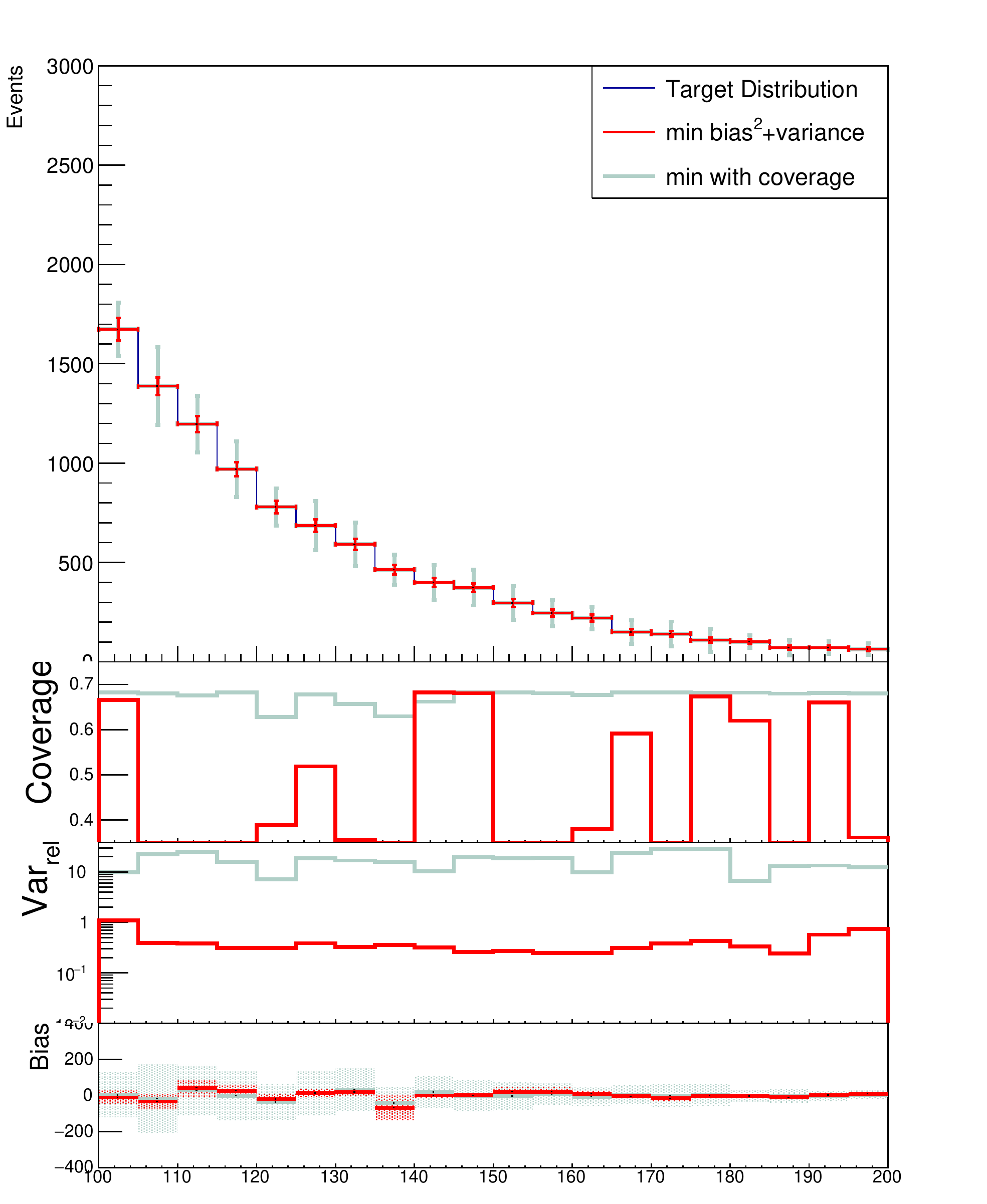} &
\includegraphics[trim=14 0 55 0,clip,width=.155\textwidth]{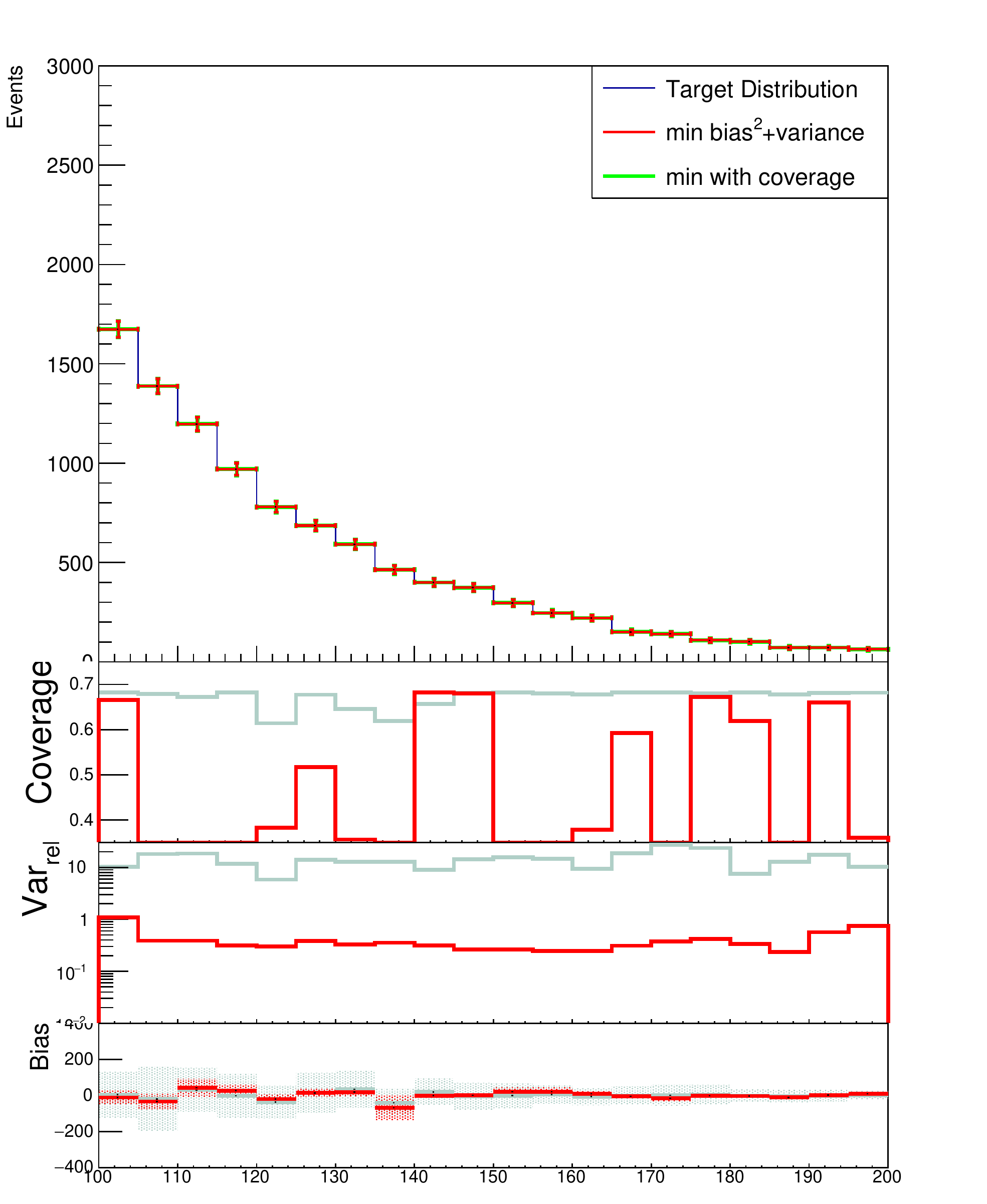} &
\includegraphics[trim=14 0 55 0,clip,width=.155\textwidth]{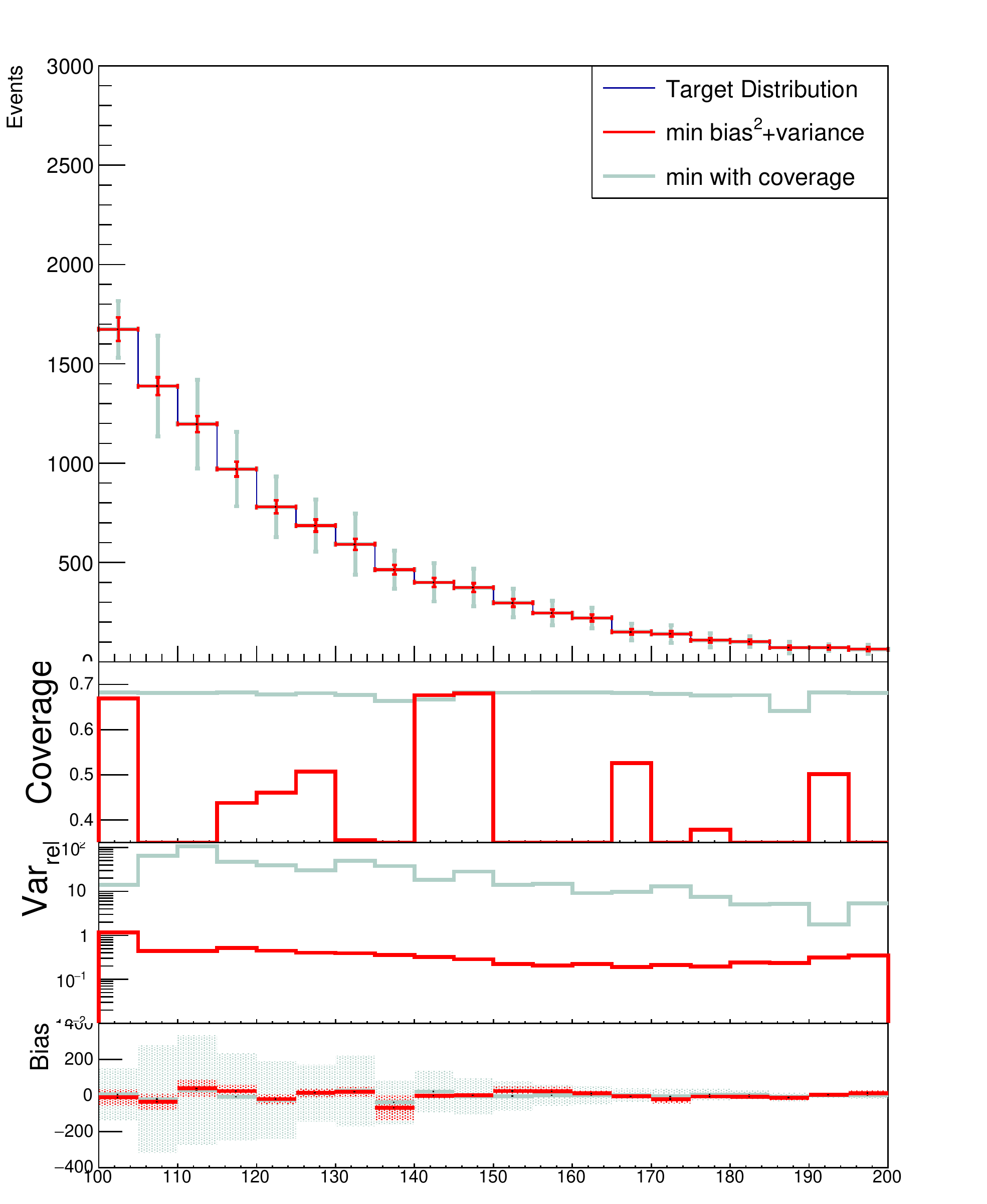} &

\includegraphics[trim=14 0 55 0,clip,width=.155\textwidth]{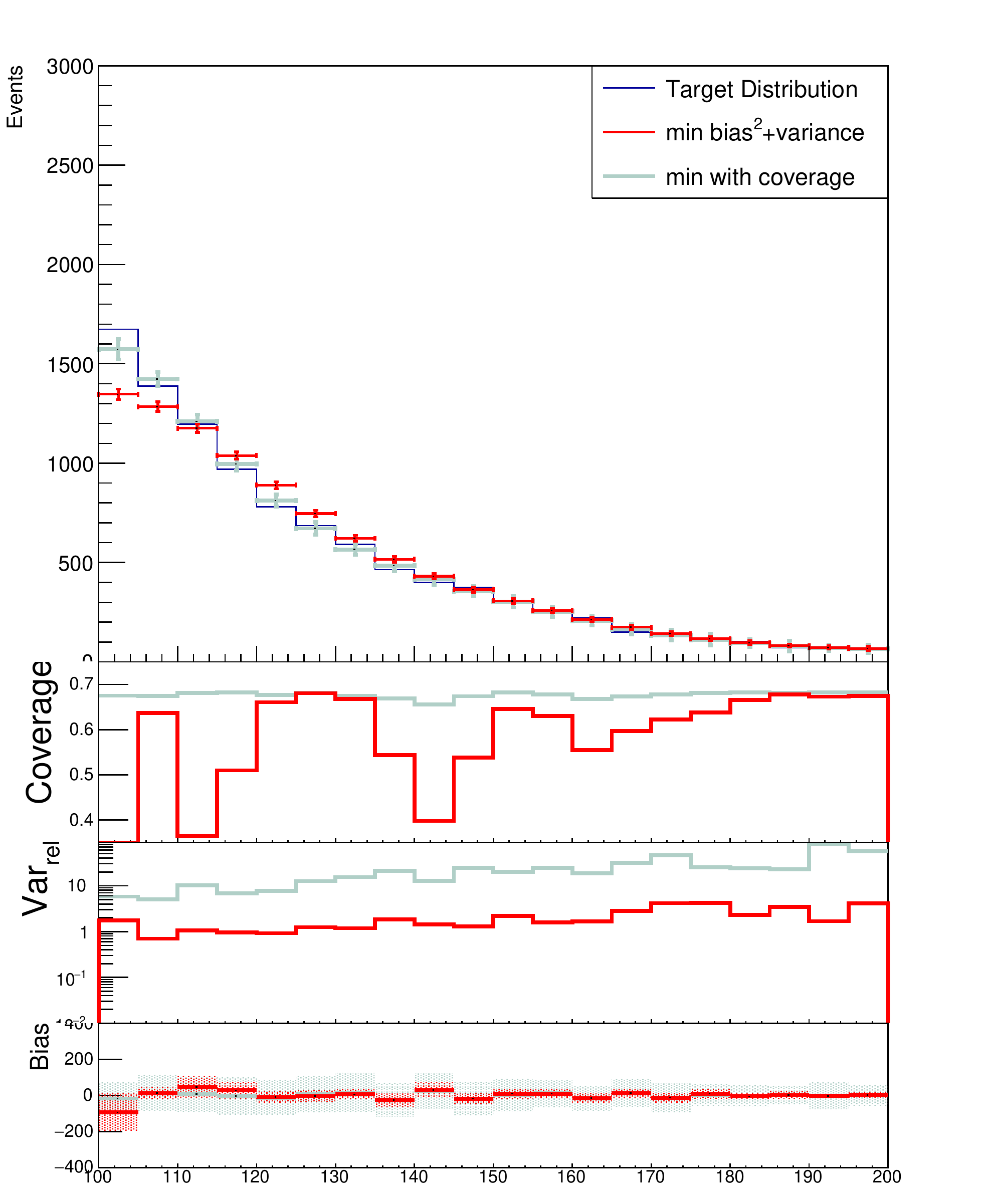} &
\includegraphics[trim=14 0 55 0,clip,width=.155\textwidth]{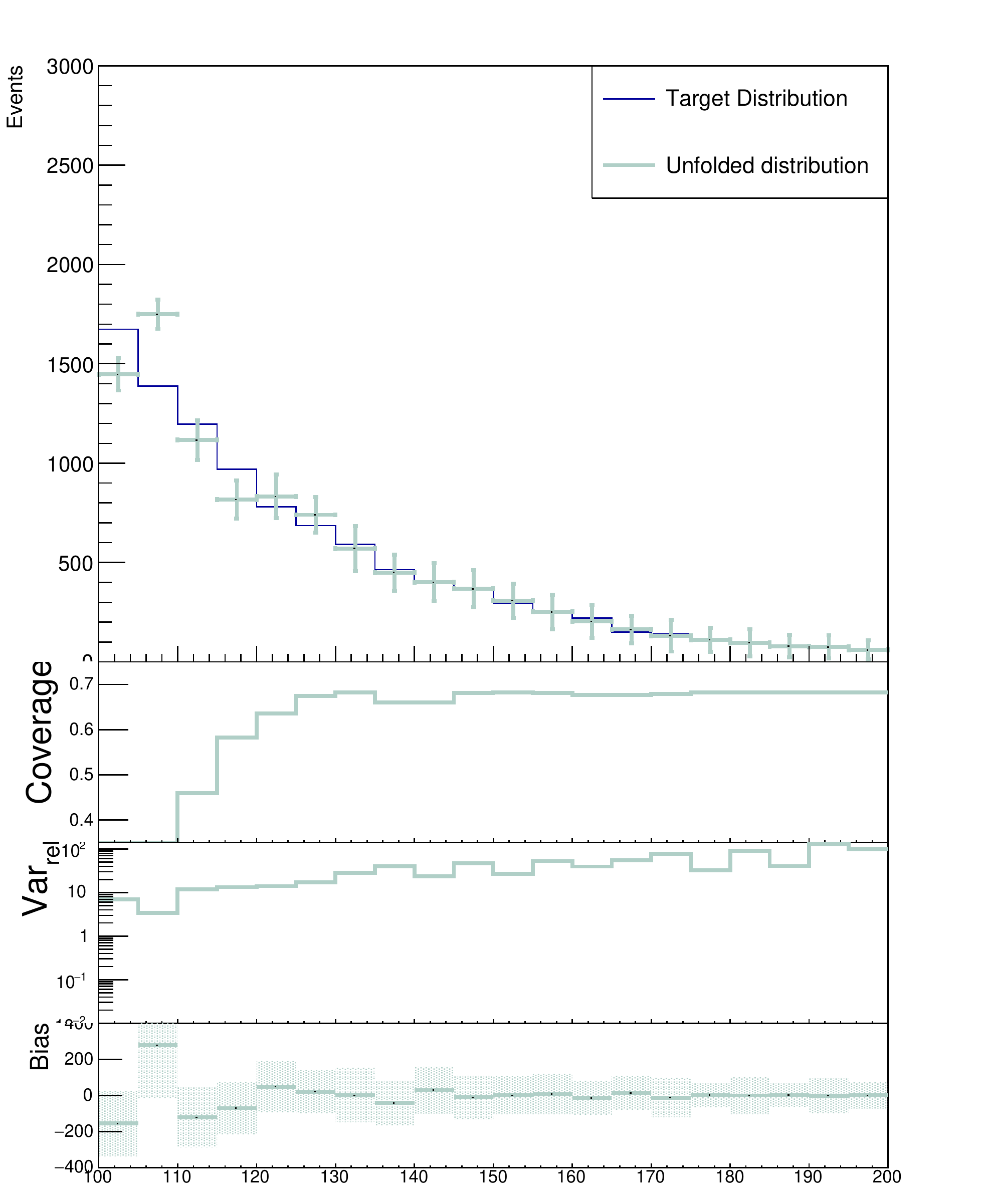} &
\includegraphics[trim=14 0 55 0,clip,width=.155\textwidth]{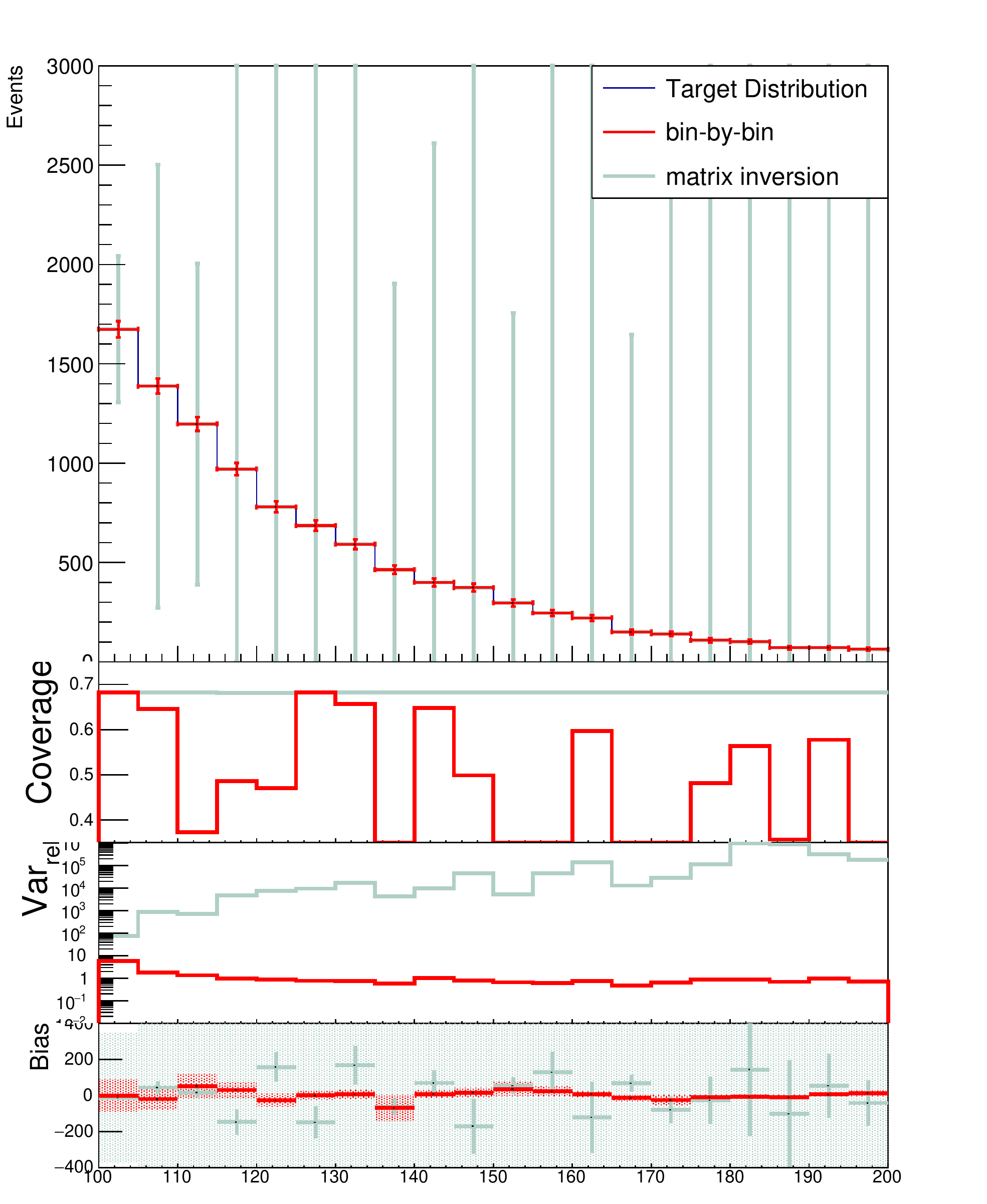} \\  
\includegraphics[trim=14 0 50 0,clip,width=.155\textwidth]{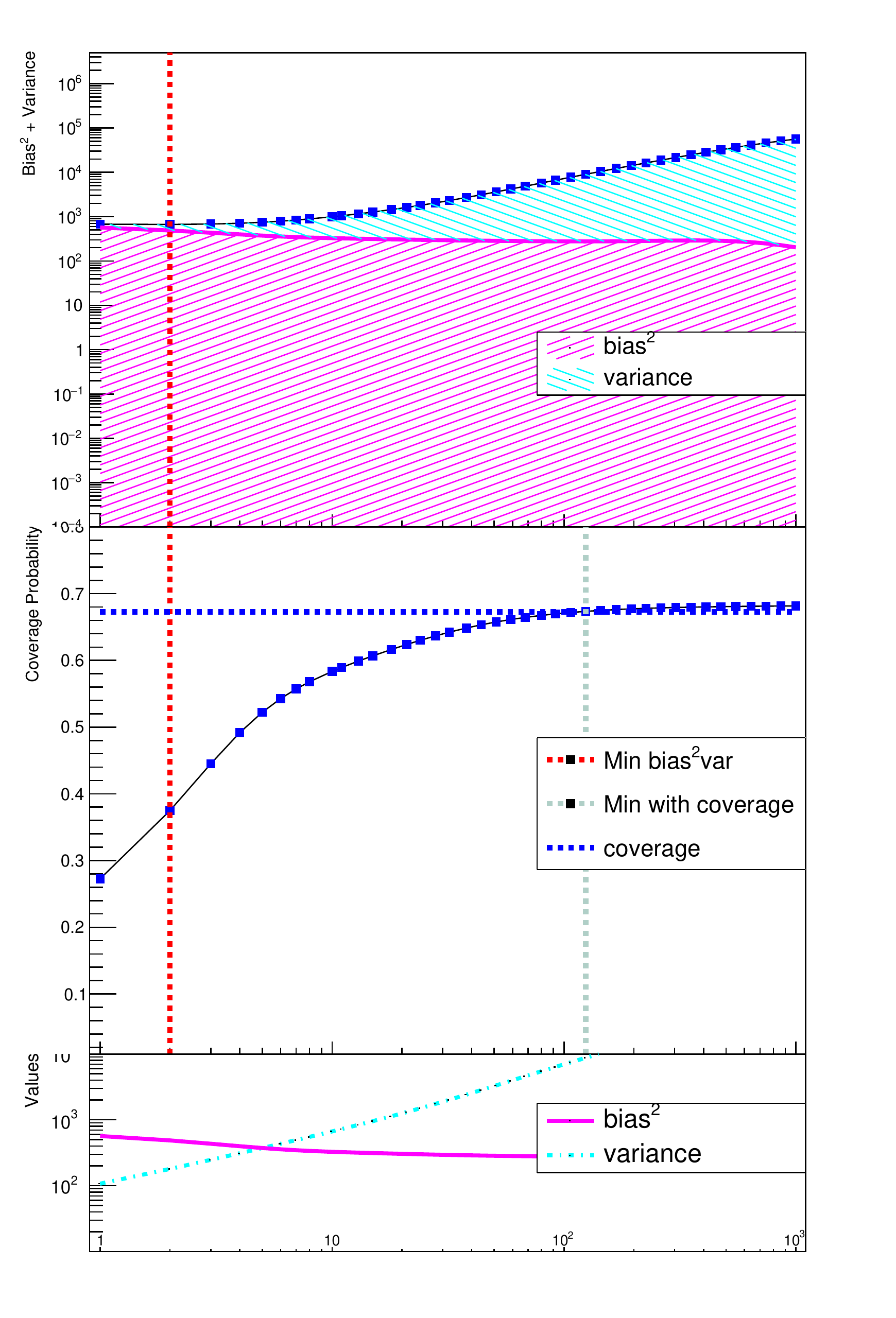} &
\includegraphics[trim=14 0 50 0,clip,width=.155\textwidth]{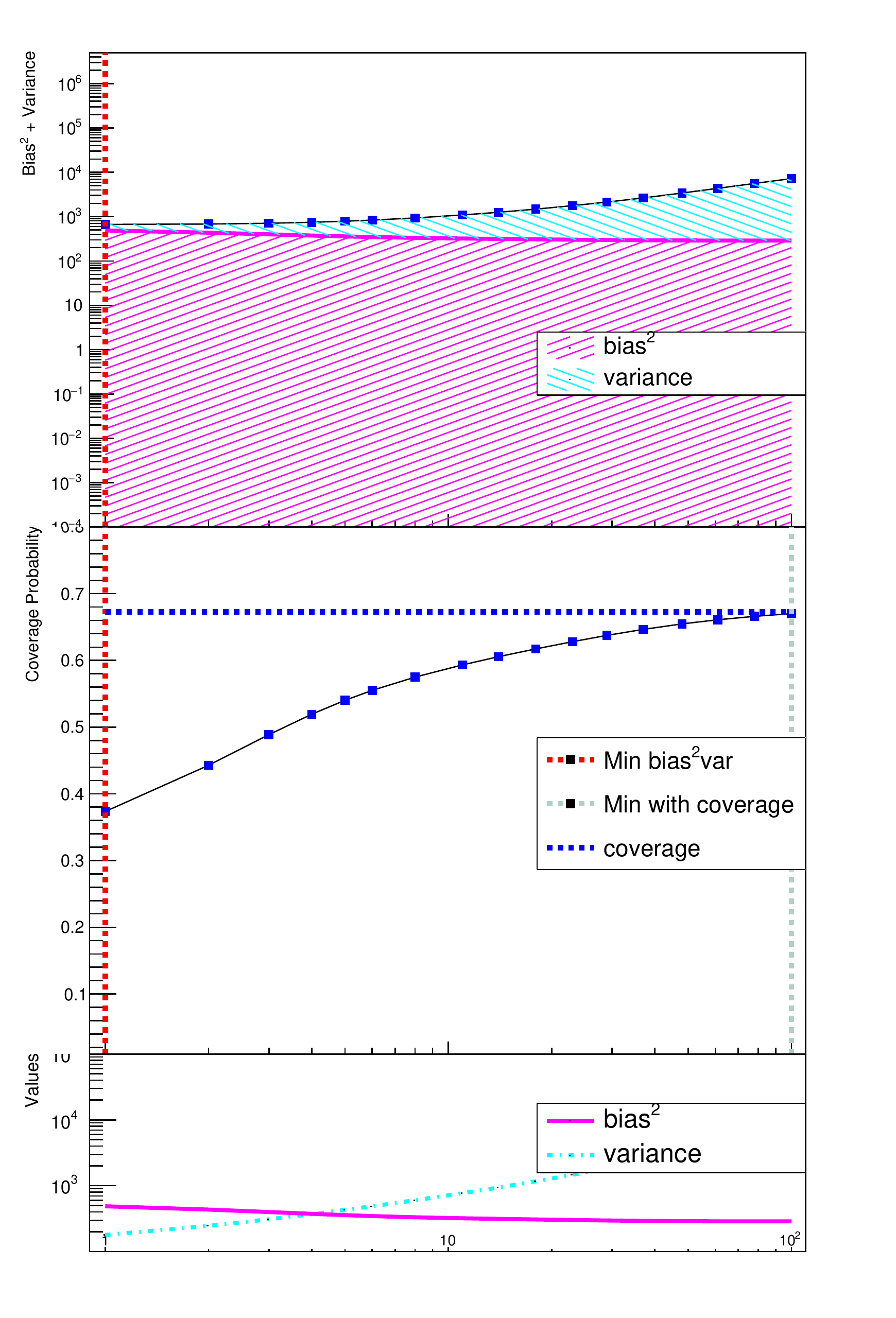} &
\includegraphics[trim=14 0 50 0,clip,width=.155\textwidth]{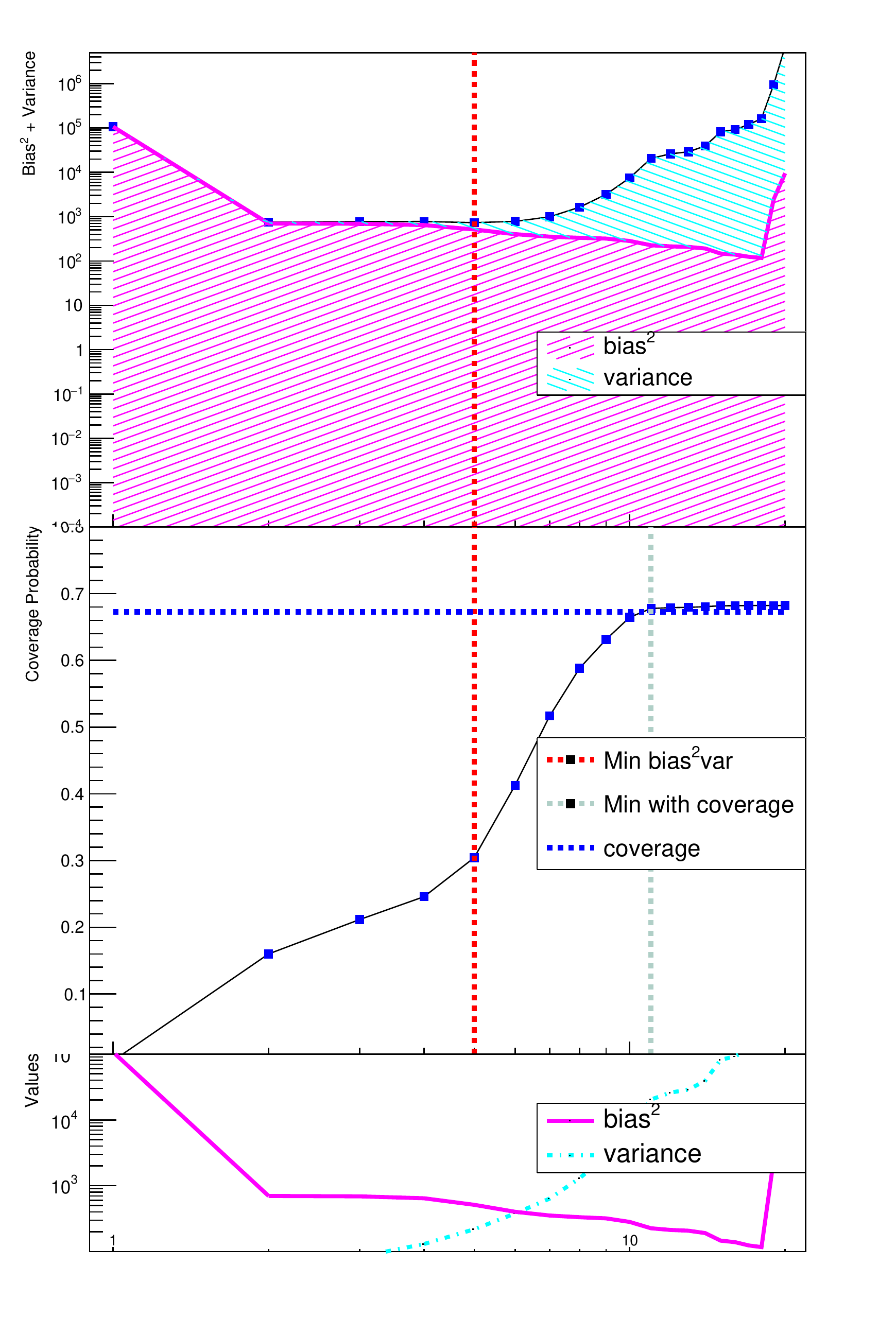} &

\includegraphics[trim=14 0 50 0,clip,width=.155\textwidth]{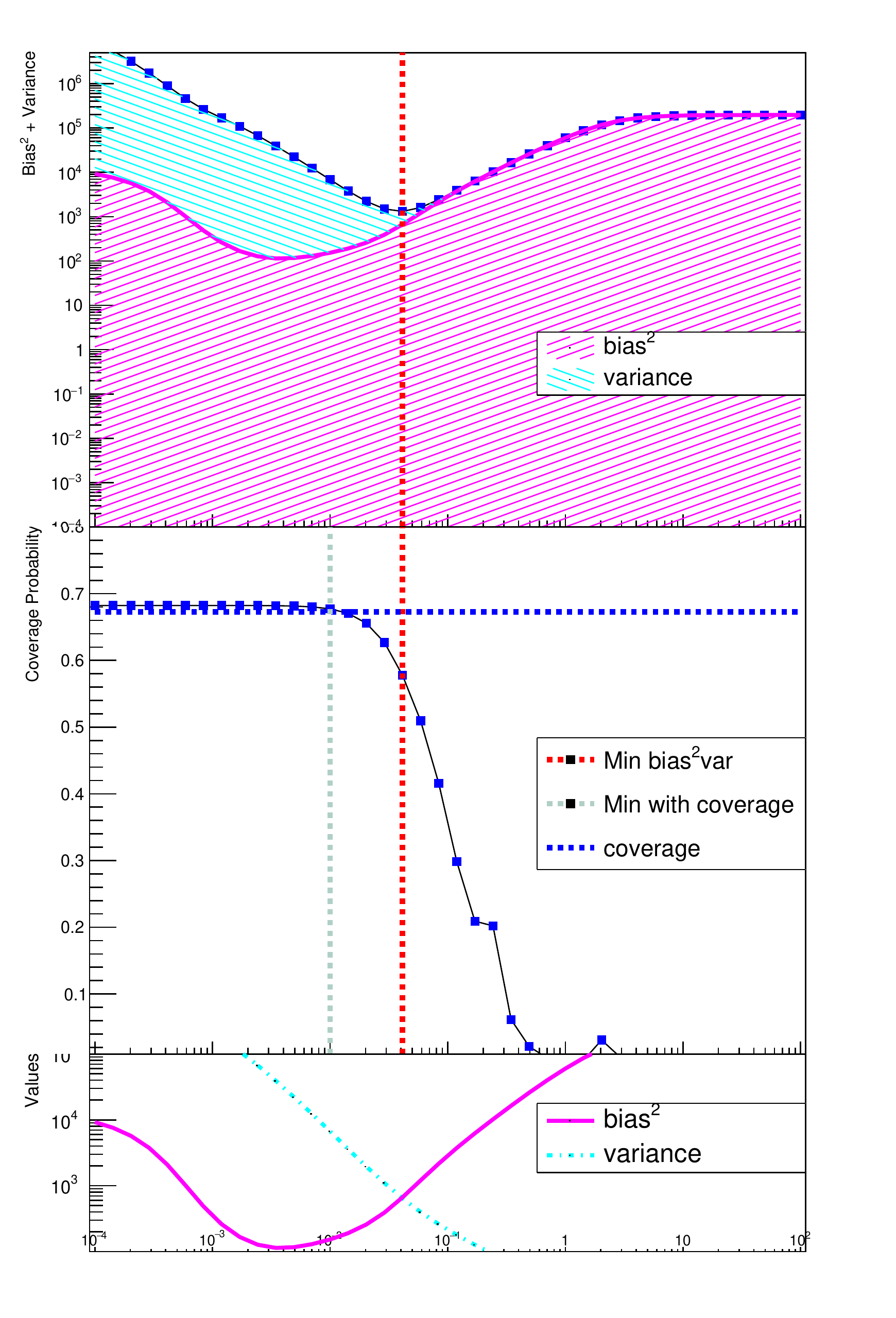} & 
\hspace{0.155\textwidth} & 
\hspace{0.155\textwidth} \\
\end{tabular} 
\caption{\label{fig:bigplot_sm}Comparison of the unfolding performance for the algorithms discussed in this paper. For each method, the top plot shows the unfolded SM data compared to the SM true
distribution, where SM response matrix was used as input to the unfolding method. For the methods with a tuneable regularisation strength two unfolding solutions are shown: (1) with a regularisation strength 
corresponding to the unconditionally minimised MSE (red), and (2) with a conditionally minimised MSE with the requirement that the bin-averaged coverage reaches the target coverage within 1\% (grey). 
The middle and lower panels of the top row plots display the bin-bin coverage, variance and bias estimates. For the bin-by-bin bias, the uncertainty of the toy-based estimate is shown with an error bar.
Also overlaid on the bias plot is the statistical error of the unfolded data (i.e the RMS of the distribution of $\hat{\mu}$) to help guide the interpretation of observed fluctuations in the unfolded data. 
For the tuneable methods, the bin-averaged MSE and the bin-averaged coverage, which are used to tune regularisation strength, are shown as function of that strength in the bottom pane. The vertical lines 
in the bottom pane indicate the regularisation strength solutions chosen by the optimisation methods (1) and (2). 
  }
\end{sidewaysfigure*}

\begin{sidewaysfigure*}[p]
\begin{tabular}{cccccc}
 & & & & & Matrix Inversion \& \\
 Bayes & 
 IDS & 
 TSVD & 
 TUnfold & 
 Gaussian Process &
Bin-by-Bin \\
\includegraphics[trim=14 0 55 0,clip,width=.155\textwidth]{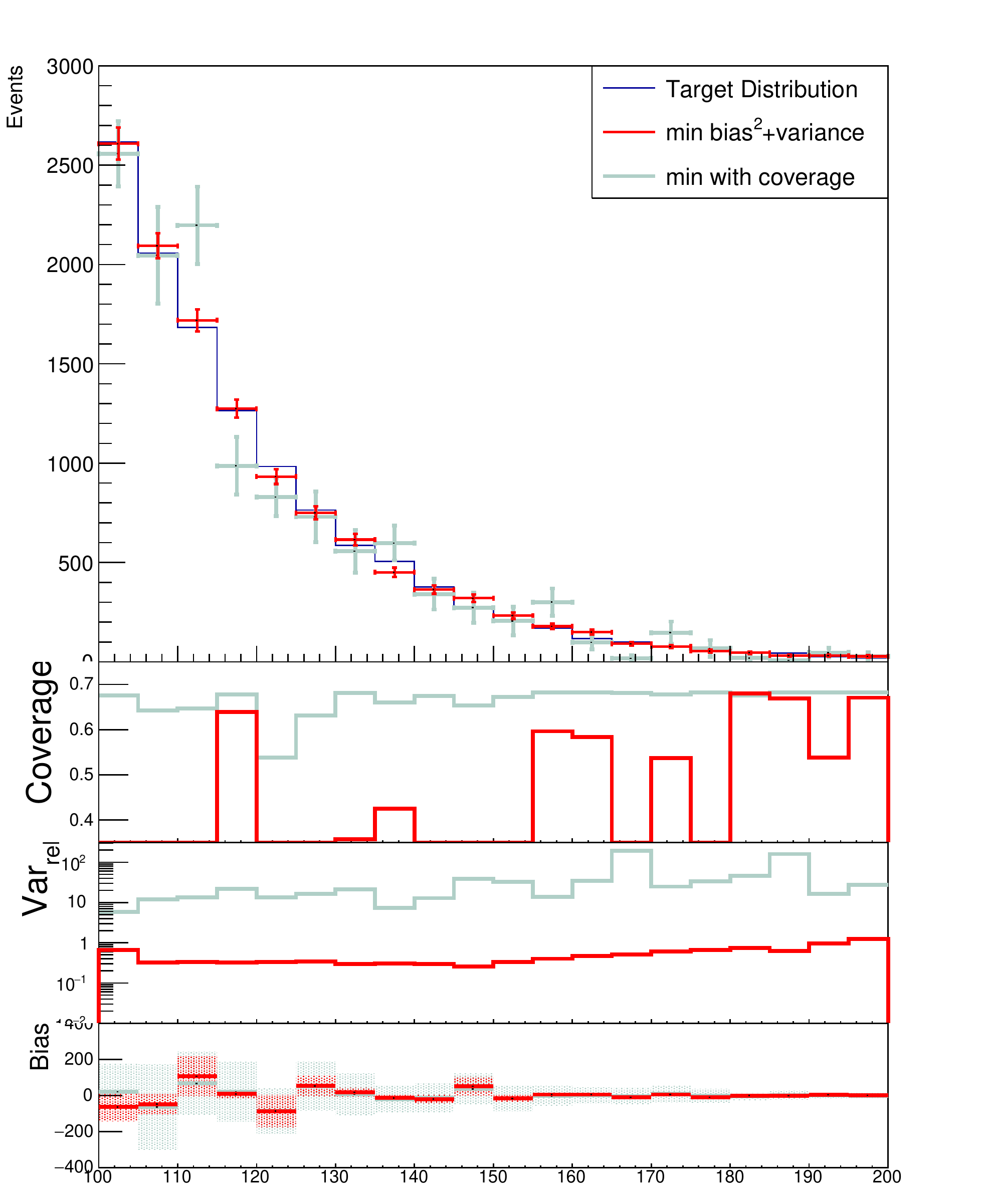} &
\includegraphics[trim=14 0 55 0,clip,width=.155\textwidth]{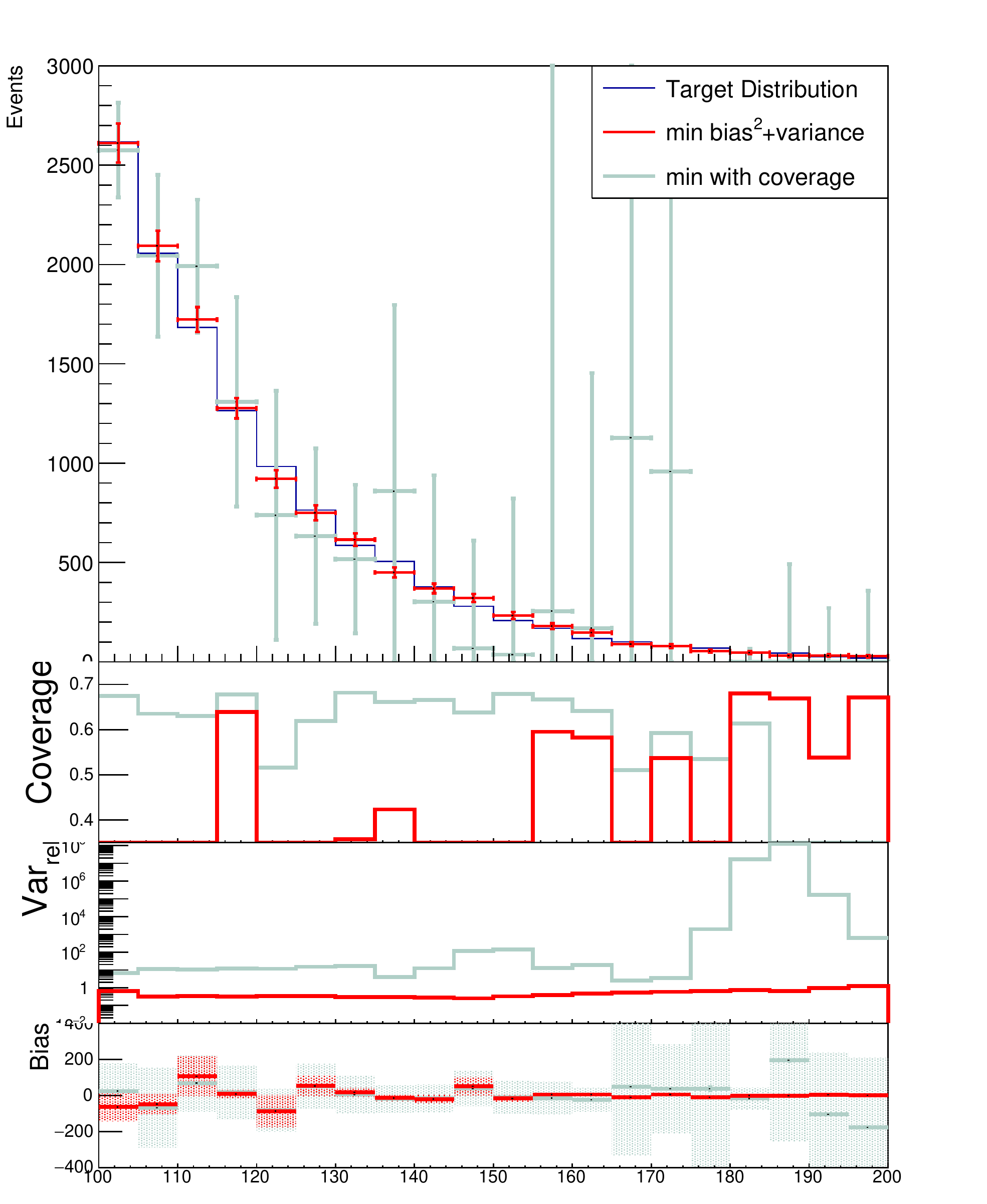} &
\includegraphics[trim=14 0 55 0,clip,width=.155\textwidth]{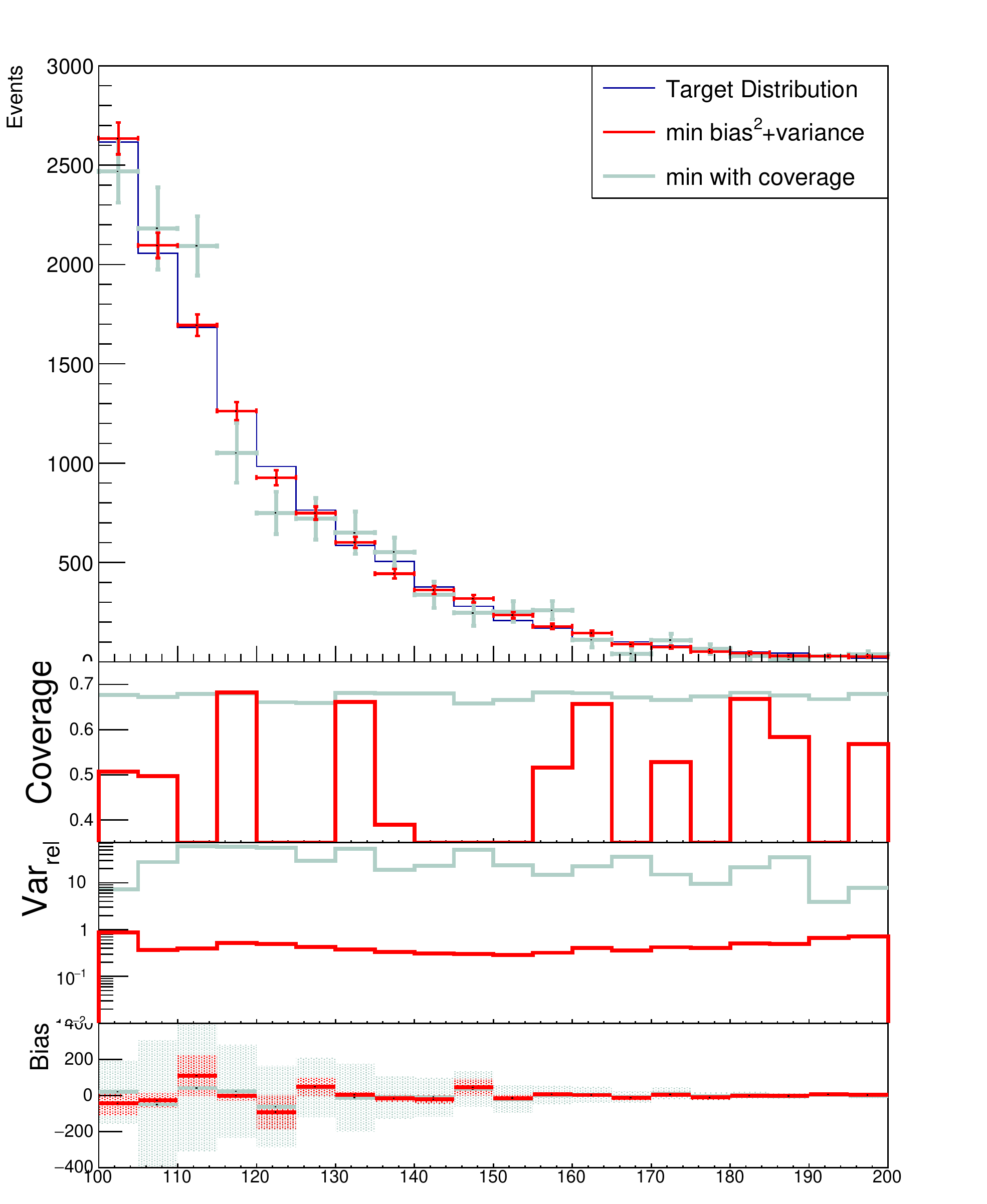} &

\includegraphics[trim=14 0 55 0,clip,width=.155\textwidth]{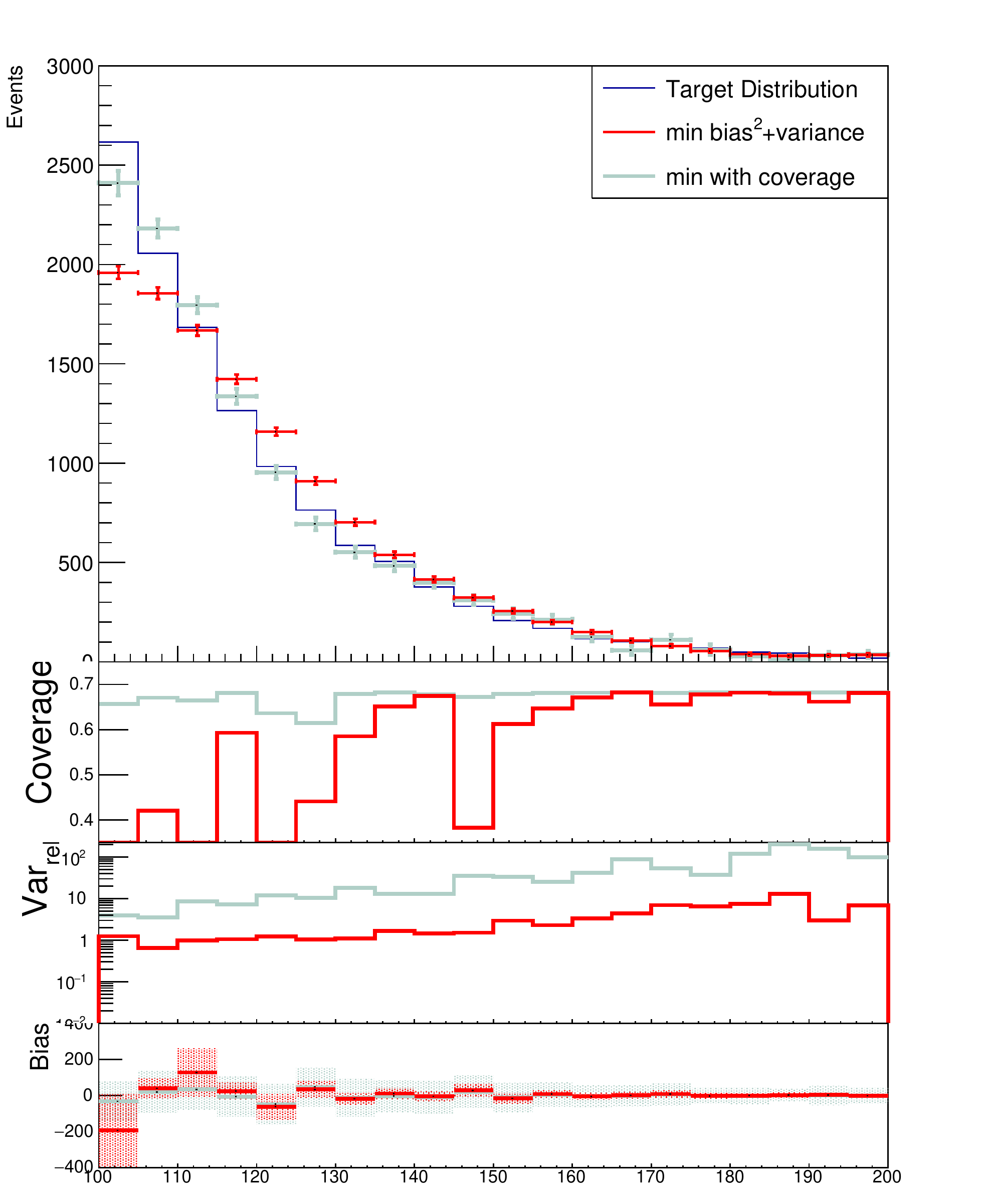} &
\includegraphics[trim=14 0 55 0,clip,width=.155\textwidth]{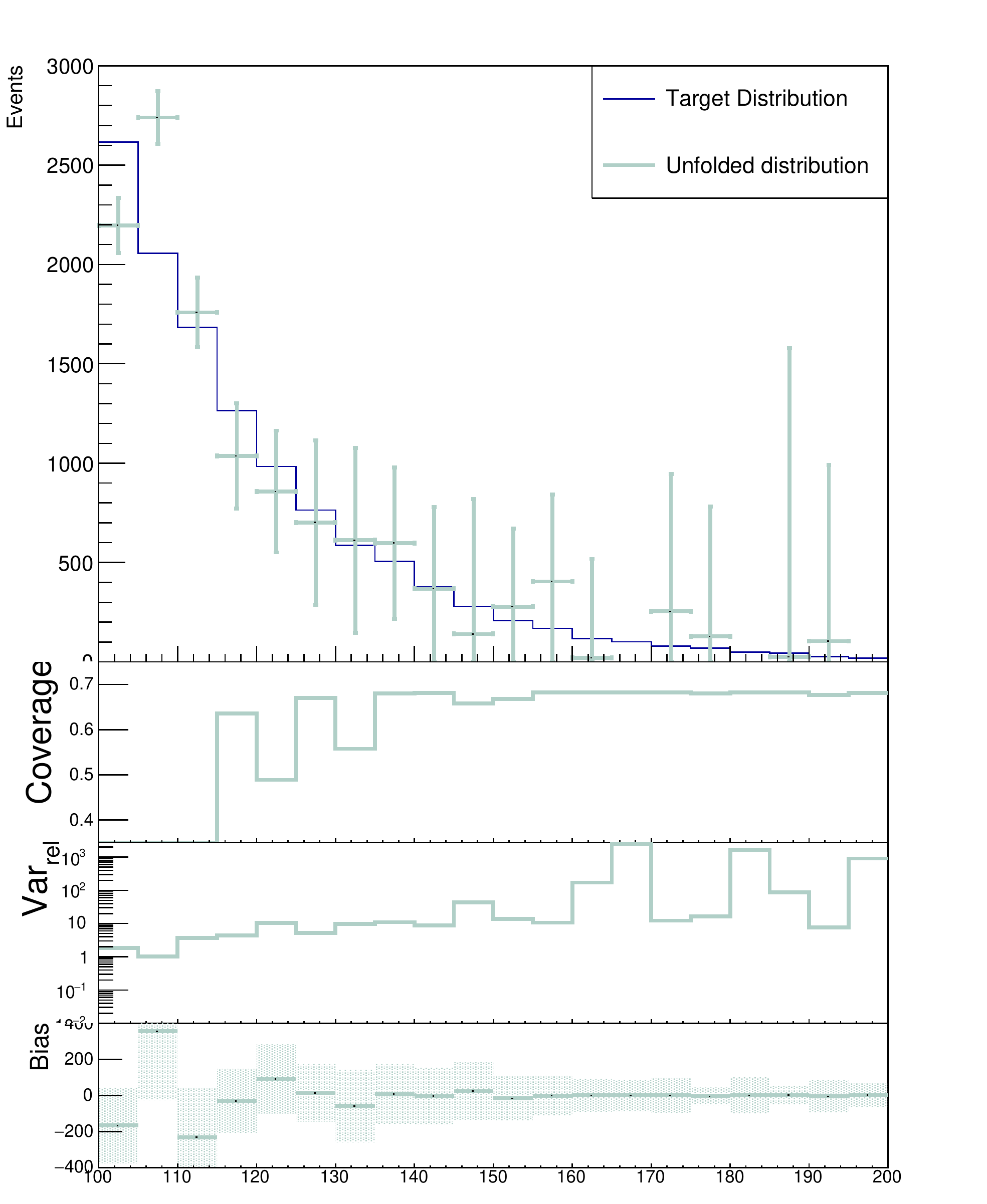} &
\includegraphics[trim=14 0 55 0,clip,width=.155\textwidth]{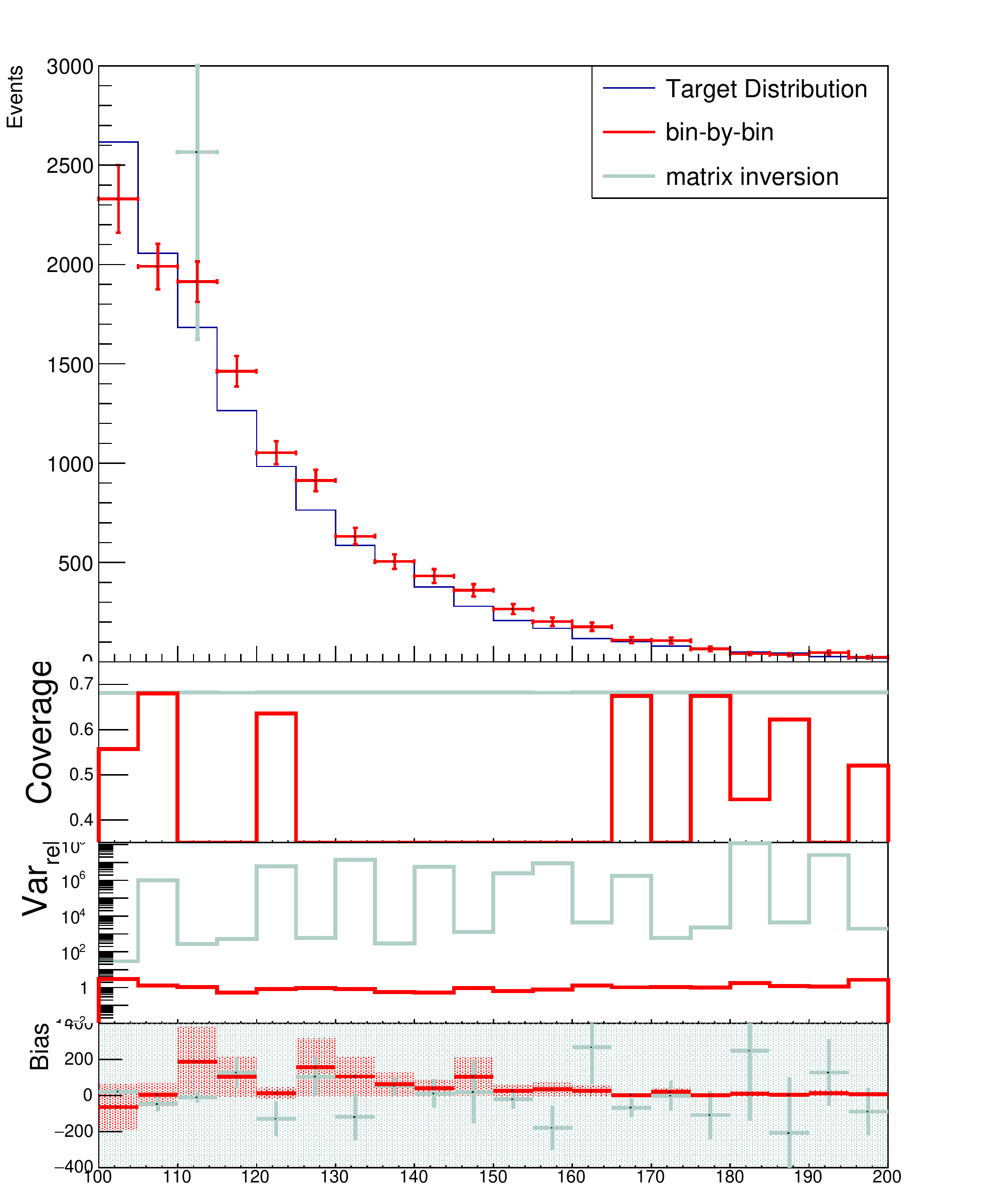} \\  
\includegraphics[trim=14 0 50 0,clip,width=.155\textwidth]{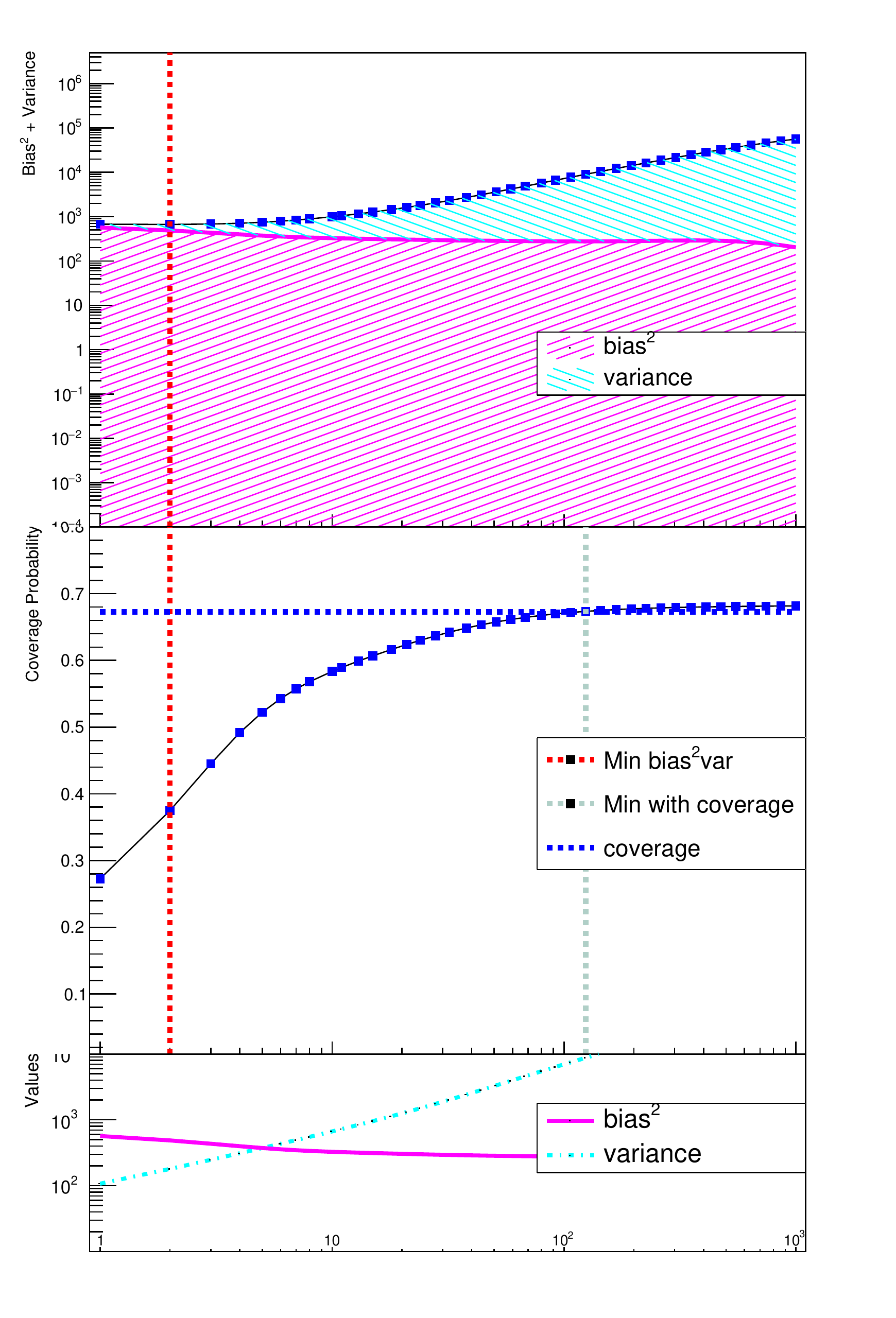} &
\includegraphics[trim=14 0 50 0,clip,width=.155\textwidth]{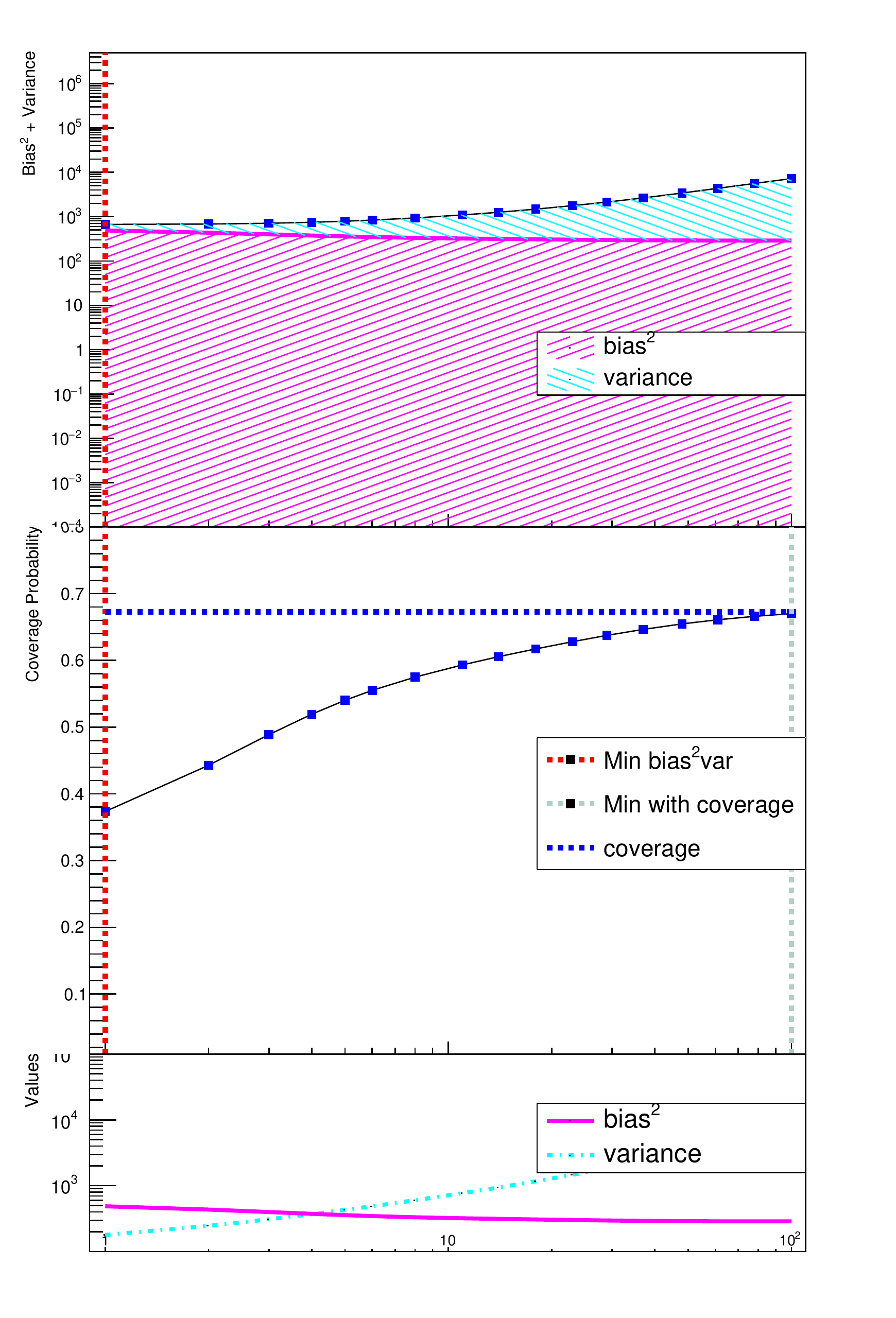} &
\includegraphics[trim=14 0 50 0,clip,width=.155\textwidth]{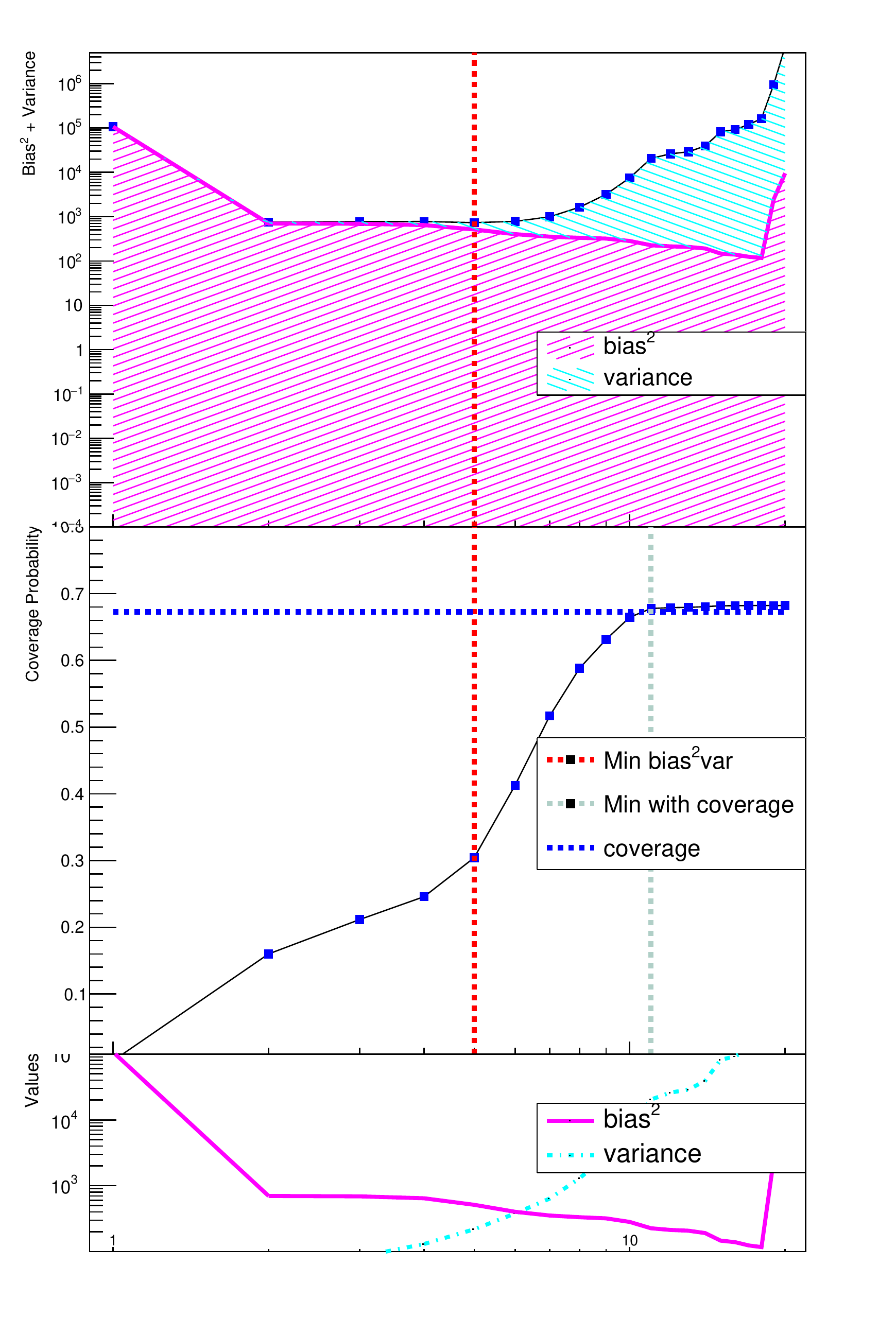} &

\includegraphics[trim=14 0 50 0,clip,width=.155\textwidth]{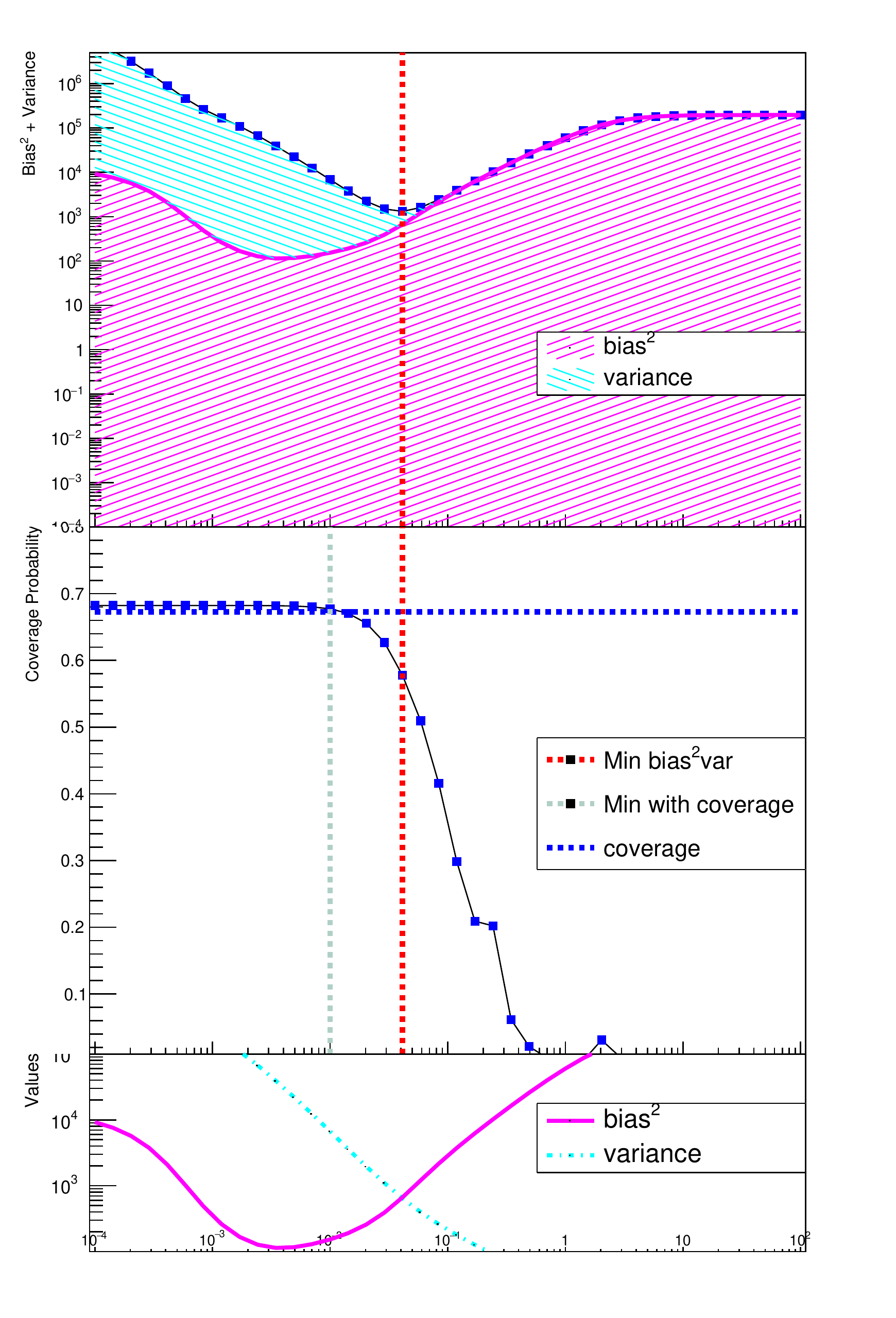} & 
\hspace{0.155\textwidth} & 
\hspace{0.155\textwidth} \\
\end{tabular} 
\caption{\label{fig:bigplot_bsm}
Comparison of the unfolding performance for the algorithms discussed in this paper. For each method, the top plot shows the unfolded BSM data compared to the BSM true
distribution, where {\em SM }response matrix was used as input to the unfolding method. For the methods with a tuneable regularisation strength two unfolding solutions are shown: (1) with a regularisation strength 
corresponding to the unconditionally minimised MSE (red), and (2) with a conditionally minimised MSE with the requirement that the bin-averaged coverage reaches the target coverage within 1\% (grey). 
In this result some of the unfolded data point fall outside the plot range, and are therefore not shown. 
The middle and lower panels of the top row plots display the bin-bin coverage, variance and bias estimates. For the bin-by-bin bias, the uncertainty of the toy-based estimate is shown with an error bar.
Also overlaid on the bias plot is the statistical error of the unfolded data (i.e the RMS of the distribution of $\hat{\mu}$) to help guide the interpretation of observed fluctuations in the unfolded data. 
For the tuneable methods, the bin-averaged MSE and the bin-averaged coverage, which are used to tune regularisation strength, are shown as function of that strength in the bottom pane. The vertical lines 
in the bottom pane indicate the regularisation strength solutions chosen by the optimisation methods (1) and (2). 
  }
\end{sidewaysfigure*}

Next, we test the unfolding algorithms in a more challenging scenario that more closely reflect the reality of unfolding problems in HEP: the observed data, corresponding to an unknown distribution (the BSM model) has to be unfolded using a response matrix obtained for a known model (the SM distribution). The benchmark in this test is how well the SM-unfolded BSM distribution compares to the BSM true distribution.  The use of the SM response matrix is a challenge for regularised methods in particular, since a too strong regularisation will bias the unfolded BSM distribution towards the SM distribution. Fig.~\ref{fig:bigplot_bsm} shows the results of this study, following  precisely the same format as that of Fig.~\ref{fig:bigplot_sm}, and demonstrates that the four tuneable unfolding methods (Iterative Bayes, IDS, TVSD, TUnfold) can also reproduce the target distribution reasonably well in this more challenging benchmark. However, in a handful of bins the bias is notably increased w.r.t the first study. The observed coverage of these four methods is qualitatively similar to that of the first study: unconditional optimisation of the regularisation strength on the bin-averaged MSE  leads to substantial undercoverage, which can again be largely 
and uniformly recovered through the imposition of a requirement on correct bin-averaged coverage in the regularisation tuning, albeit with a bit more variability in bin-by-bin coverage. For the Gaussian Process unfolding method the
main difference with the SM benchmark result is the striking increase in variance for the low statistics bins in the region $x>120$. A similarly striking increase in variance is also observed for the Matrix Inversion method, whereas the degradation of the
performance of the Bin-by-Bin method is comparable to that of the tuneable algorithms. 

\clearpage
\subsection{Bimodal example}
\label{sec:bimodal}

\begin{figure}[hbt]
\begin{tabular}{ccc}
 $\alpha=0.5$ & $\alpha=1$ & $\alpha=1.5$ \\
\includegraphics[trim=0 0 0 0,clip,width=0.32\textwidth]{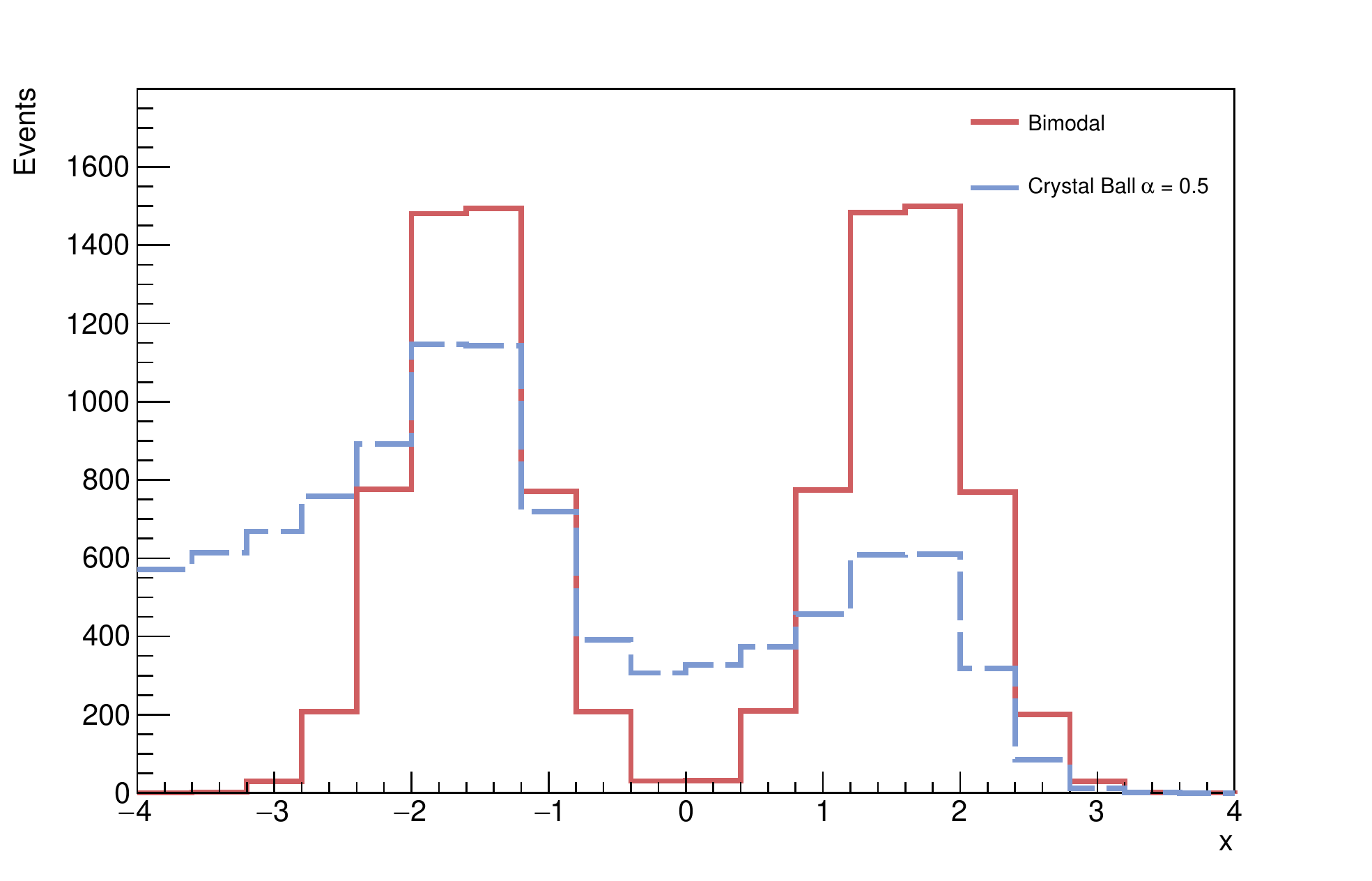} &
\includegraphics[trim=0 0 0 0,clip,width=0.32\textwidth]{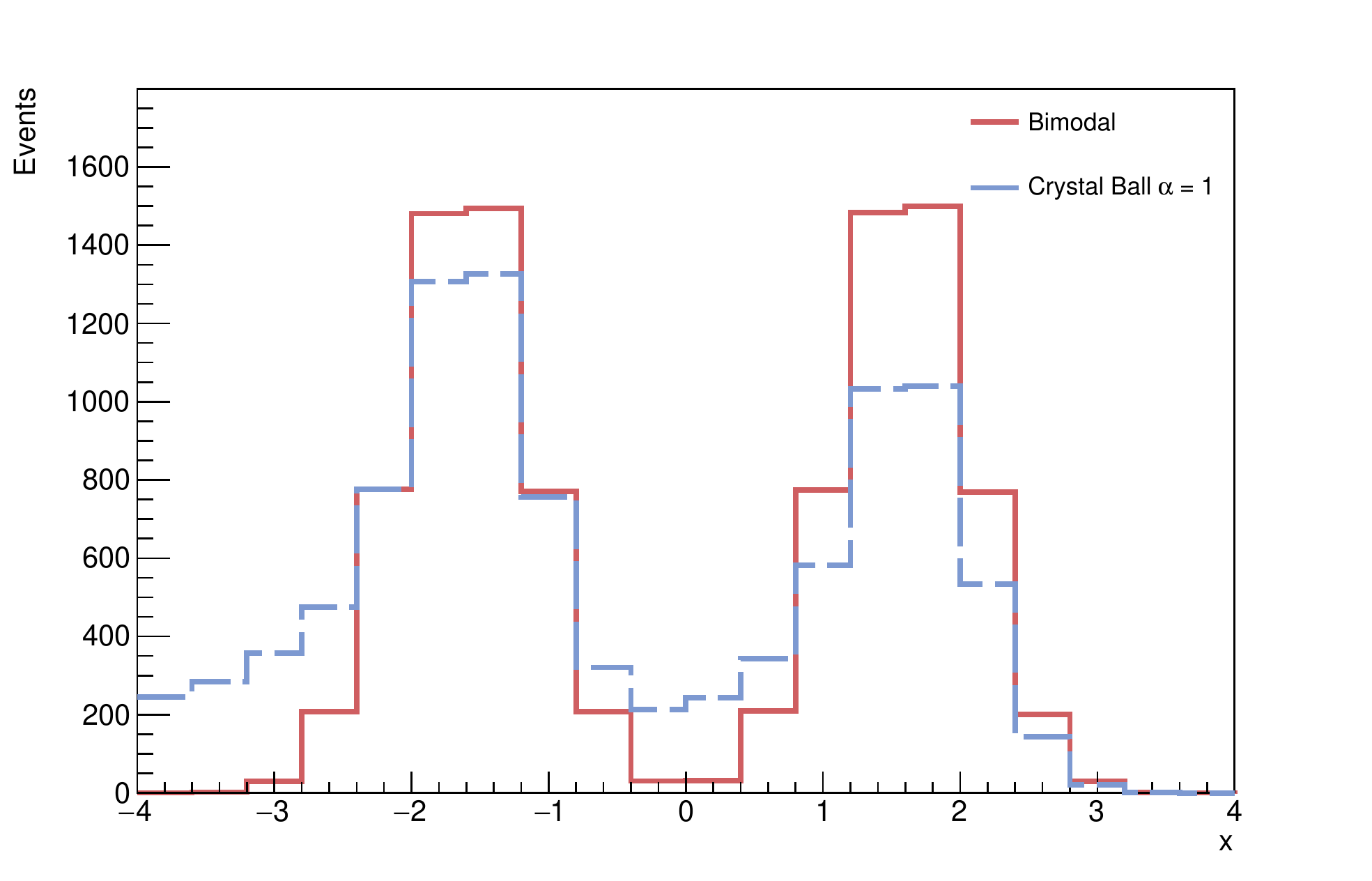} &
\includegraphics[trim=0 0 0 0,clip,width=0.32\textwidth]{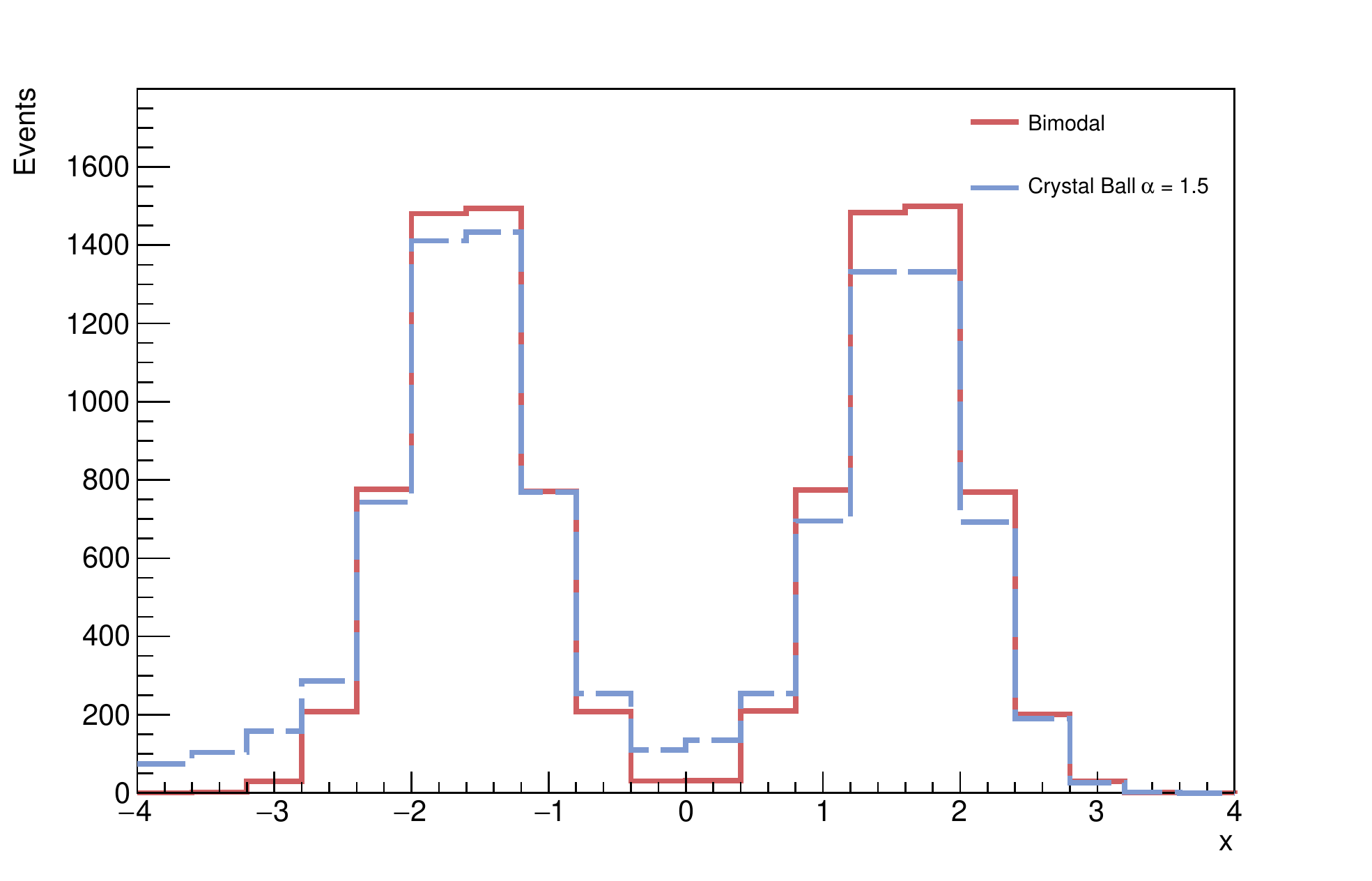} \\
$\alpha=2.0$ & $\alpha=2.5$ & $\alpha=3.0$ \\
\includegraphics[trim=0 0 0 0,clip,width=0.32\textwidth]{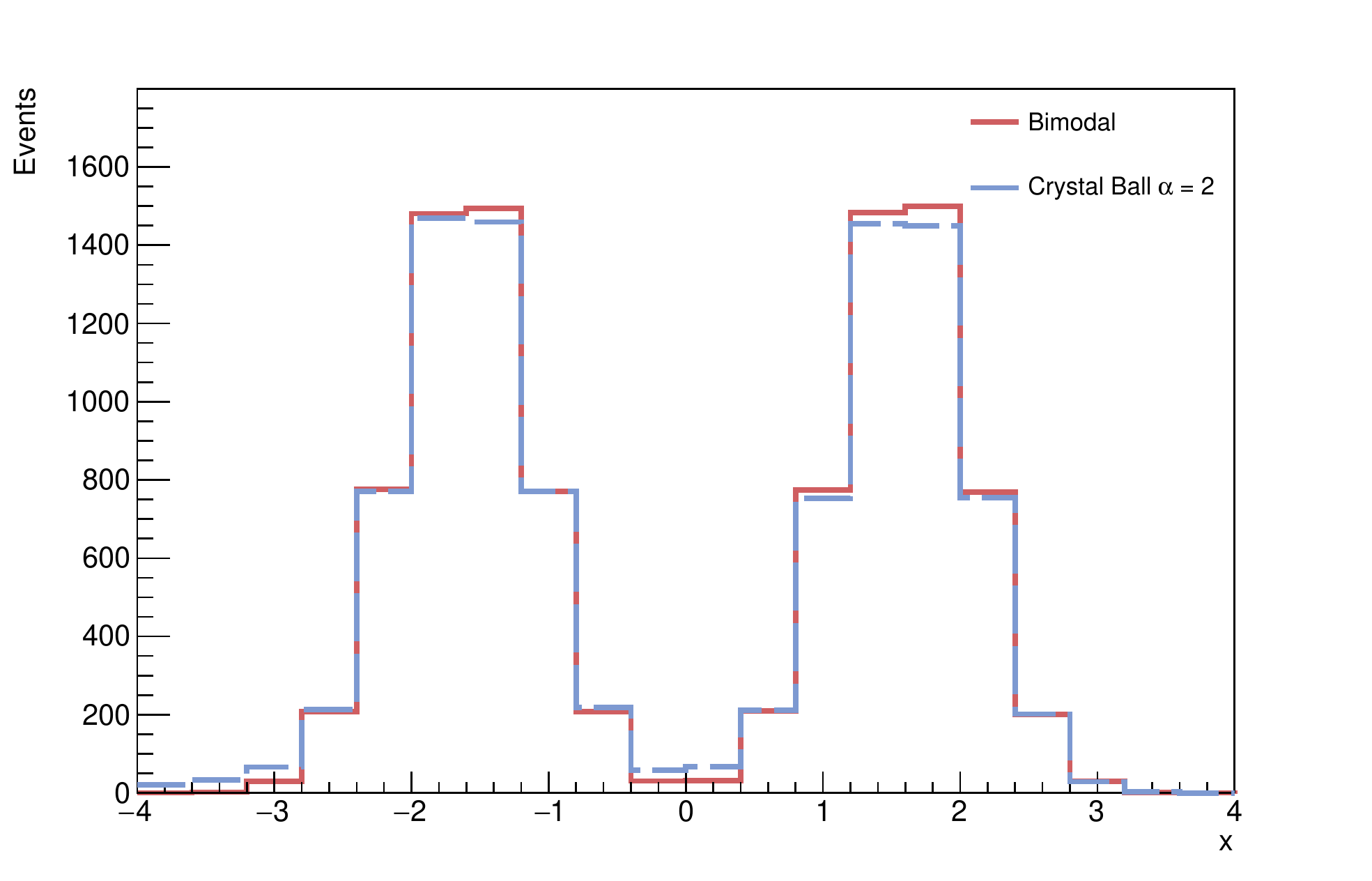} &
\includegraphics[trim=0 0 0 0,clip,width=0.32\textwidth]{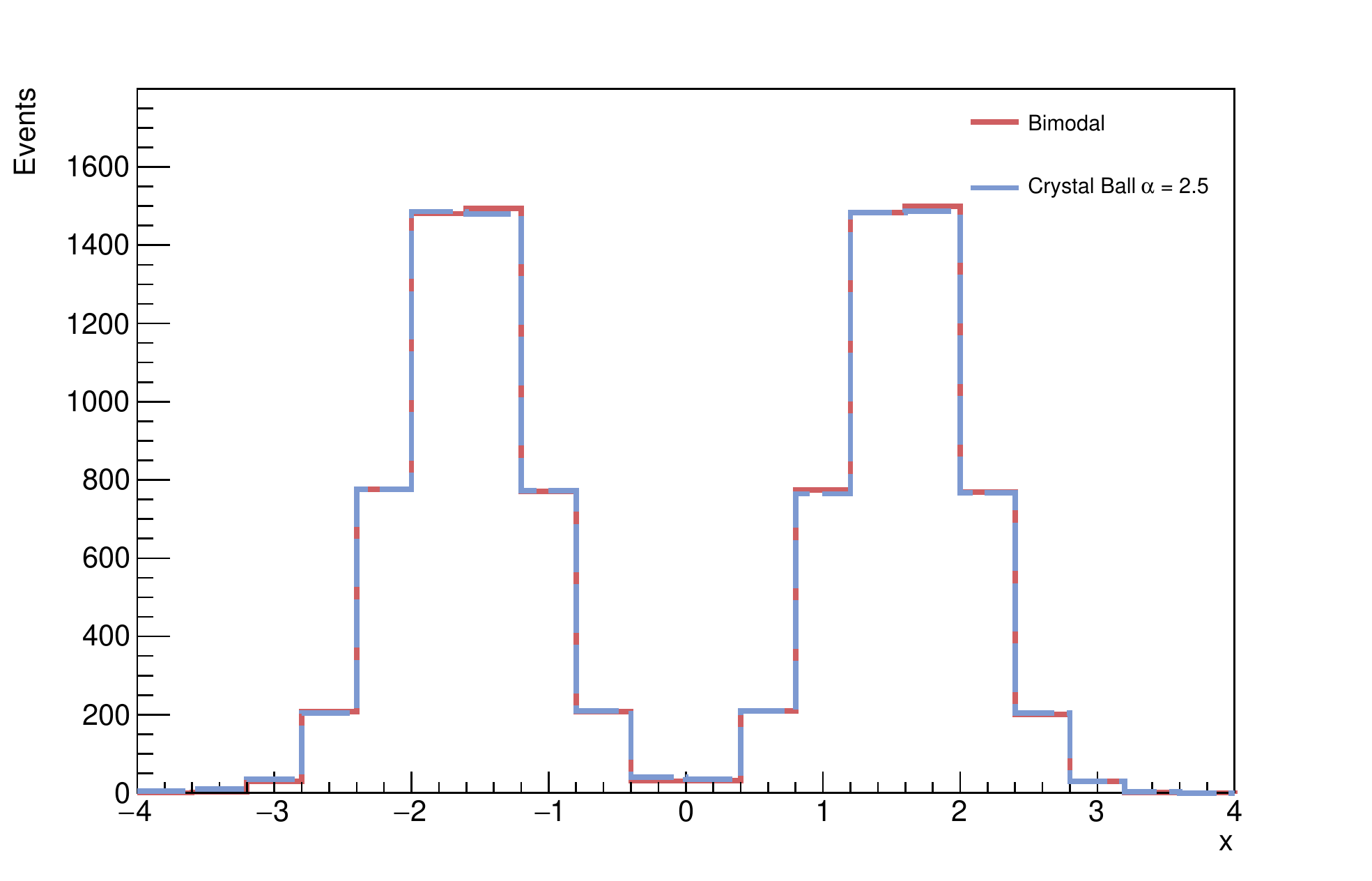} &
\includegraphics[trim=0 0 0 0,clip,width=0.32\textwidth]{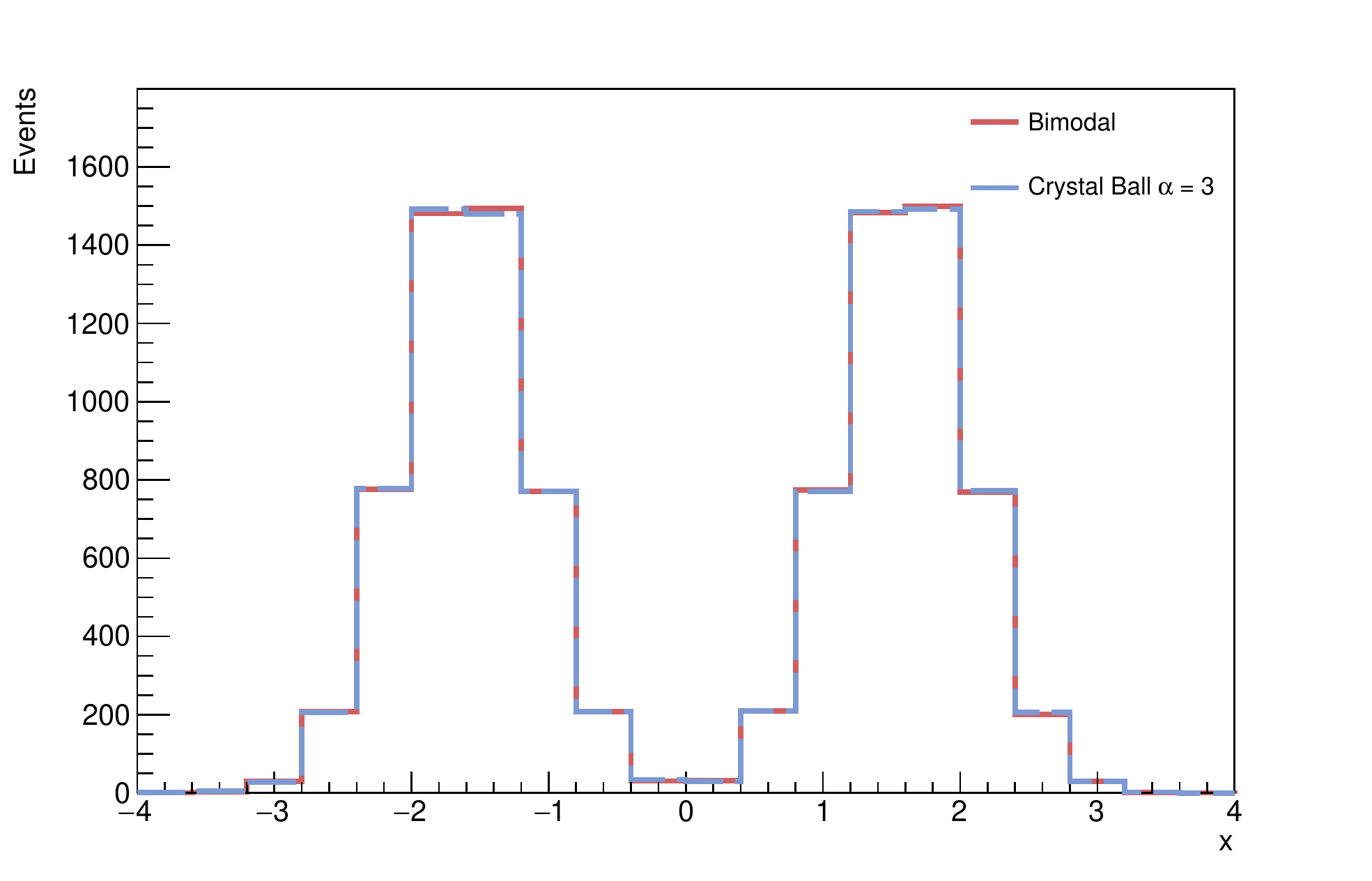} \\
\end{tabular} 
\caption{\label{fig:scan_bimodal} Visualization of the true distribution of the bimodal model for various values of the distortion parameter $\alpha$. }
\end{figure}

We use a bimodal distribution to further investigate unfolding biases in situations where the response matrix and the data are not sampled from the same model.
The bimodal model for this study is the the sum of two Crystal Ball~\cite{Oreglia:1980cs} functions, which is smeared by a simple Gaussian resolution model

\begin{eqnarray*}
f(x|\alpha) & = & f_\textrm{physics}(x_\textrm{true}|\alpha) \ast f_\textrm{detector}(x_\textrm{true},x) \\
                  & =  & ( 0.5 \cdot f_{CB}(x_\textrm{true}|\mu=2.4,\sigma=0.48,\alpha,n=1) +  \\
                  &     &  \: 0.5 \cdot  f_{CB}(x_\textrm{true}|\mu=5.6,\sigma=0.48,\alpha,n=1)  ) \\
                  &     & \ast \; Gauss(x-x_\textrm{true},0,0.4)
\end{eqnarray*} 

\noindent where the Crystal Ball probability density function $f_{CB}(x|\mu,\sigma,\alpha,n)$ has a Gaussian core with mean $\mu$ and width $\sigma$, and a power law tail with power $n$ below a threshold $\alpha$ that is expressed units
of the Gaussian width $\sigma$. Thus, for $\alpha=\infty$ a Crystal Ball function is identical to a Gaussian distribution, whereas e.g. for $n=1$ it follows a Gaussian
distribution for $x>-1 \sigma$ and a power law distribution  $x \le -1\sigma$. In the definition of the Crystal Ball function the normalisation factor for the 
power law tail is chosen such that the function is continuous and differentiable over the transition point. The true and expected data distributions $\vec{\mu}$ and $\vec{\nu}$ 
corresponding to this model are defined by 20 uniformly sized bins in the range $[-4,4]$ and are populated with 10000 events. Fig.~\ref{fig:scan_bimodal} shows the true distribution of the bimodal
model for a range $\alpha$ values.

\begin{figure}[hbt]
\begin{center}
\includegraphics[trim=38 0 50 30,clip,width=0.70\textwidth]{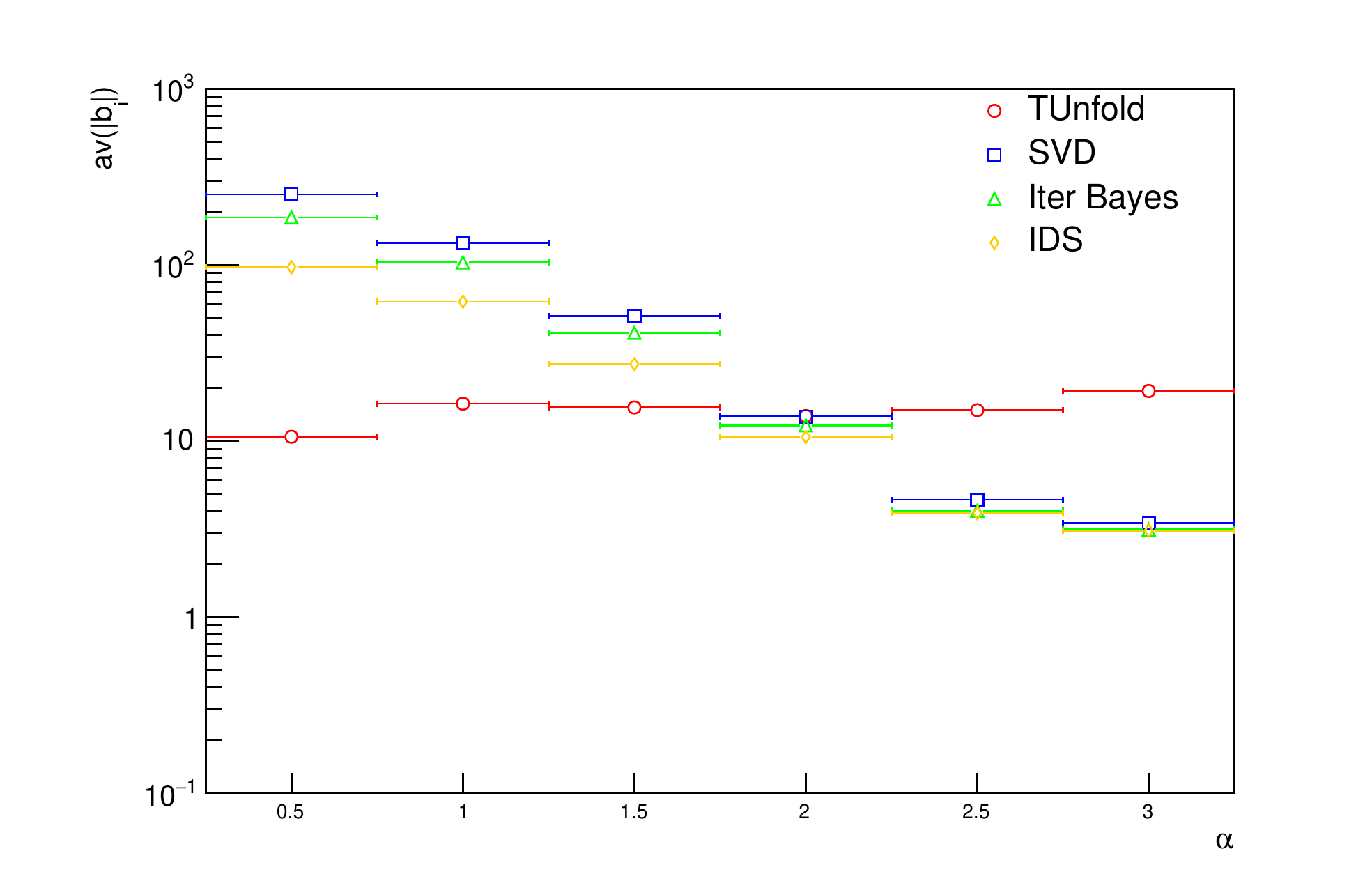} 
\caption{\label{fig:bias_bimodal_mse} Comparison of the estimated bin-averaged bias of the  tuneable unfolding algorithms 
when unfolding the distorted bimodal model with a response matrix obtained from the undistorted model. Results are shown for various distortion
strengths, which increase in distortion with decreasing values of $\alpha$ (See Fig.~\protect{\ref{fig:scan_bimodal}}). In this study, the regularisation strength for each algorithm
has been unconditionally optimised on the MSE.}
\end{center}
\end{figure}

Fig.~\ref{fig:bias_bimodal_mse} shows the bin-averaged unfolding bias of the tuneable algorithms (Iterative Bayes, IDS, SVD, TUnfold) for data sampled from the bimodal distribution with $\alpha=[ 0.5,1,1.5,2,2.5,3 ]$ and unfolded with a 
response matrix with $\alpha=\infty$, with the regularisation strength always tuned to the unconditionally minimised MSE. Fig.~\ref{fig:bias_bimodal_mse} shows that the SVD, Iterative Bayes and IDS methods 
exhibit a strong unfolding bias that increases with the distortion for values of $\alpha<2.5$.  In contrast, the bin-averaged bias of TUnfold is almost independent on the model distortion, which may be explained
by the fact that TUnfold regularises on smoothness, rather than the true distribution. As a result, TUnfold has the smallest bin-averaged bias for the strongest
distortions ($\alpha=0.5,1,1.5$), whereas SVD, Iterative Bayes and IDS have the (almost identical) best performance at low distortion ($\alpha=2.5,3$). 

\begin{figure}[hbt]
\begin{center}
\includegraphics[trim=38 0 50 30,clip,width=0.70\textwidth]{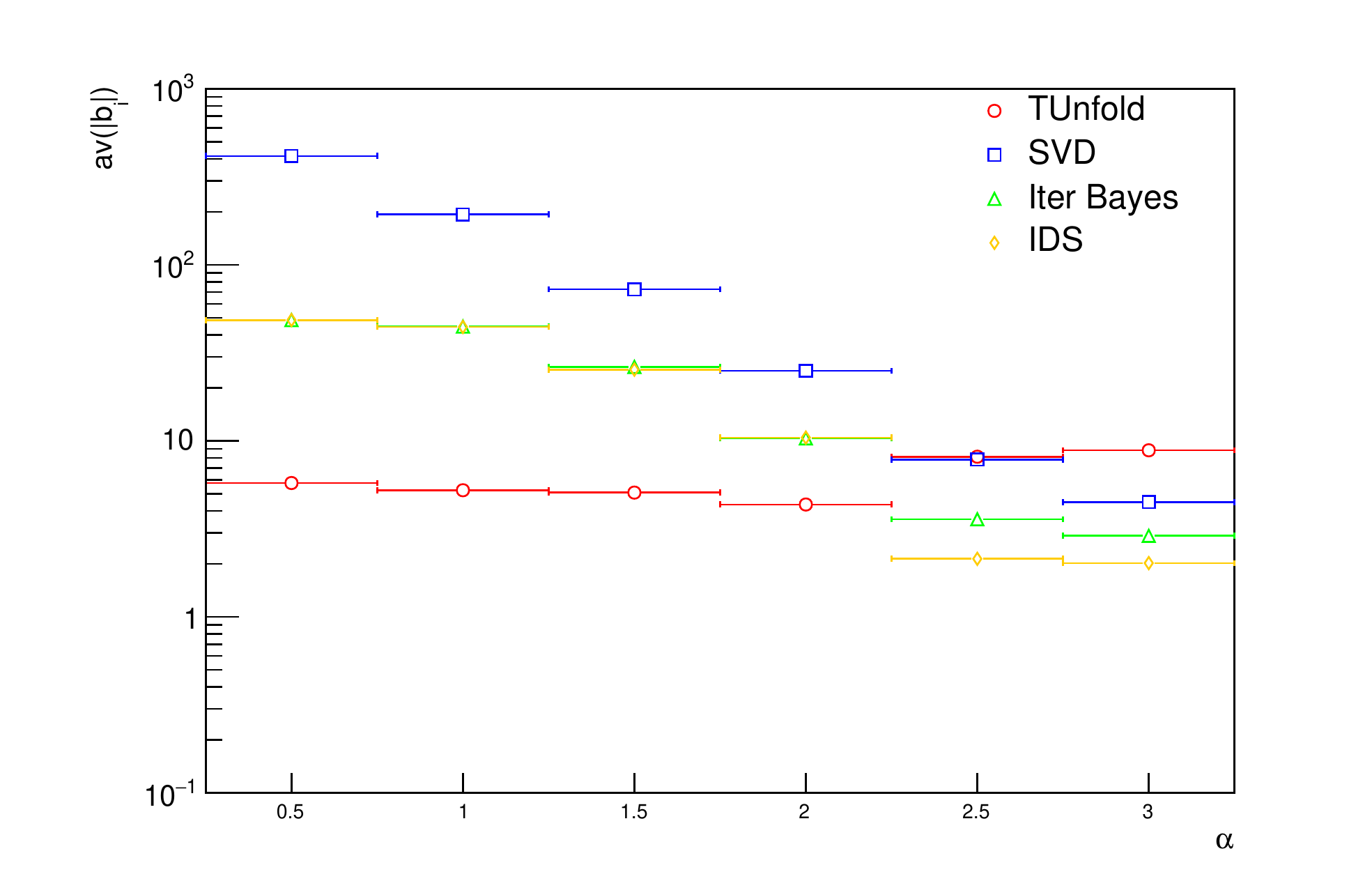} 
\caption{\label{fig:bias_bimodal_cov}  Comparison of the estimated bin-averaged bias of the  tuneable unfolding algorithms
when unfolding the distorted bimodal model with a response matrix obtained from the undistorted model. Results are shown for various distortion
strengths, which increase in distortion with decreasing values of $\alpha$ (See Fig.~\protect{\ref{fig:scan_bimodal}}). In this study, the regularisation strength for each algorithm
has been optimised on the MSE with the condition that coverage is achieved with 1\% of the target value}
\end{center}
\end{figure}

 Fig.~\ref{fig:bias_bimodal_cov} shows the same comparison as Fig.~\ref{fig:bias_bimodal_mse}, but with the average coverage condition imposed the regularisation optimisation. 
 The dependence of average bias on the distortion is qualitatively the same for all methods, with TUnfold again being the least sensitive to distortions.
 In terms of absolute performance, TUnfold gives the smallest average bias for  ($\alpha=0.5,1,1.5,2.0$) for medium to high distortion, where IDS
 outperforms for small distortions ($\alpha=2.5,3.0$). 

\begin{figure}[hbt]
\begin{center}
\includegraphics[trim=38 0 50 30,clip,width=0.70\textwidth]{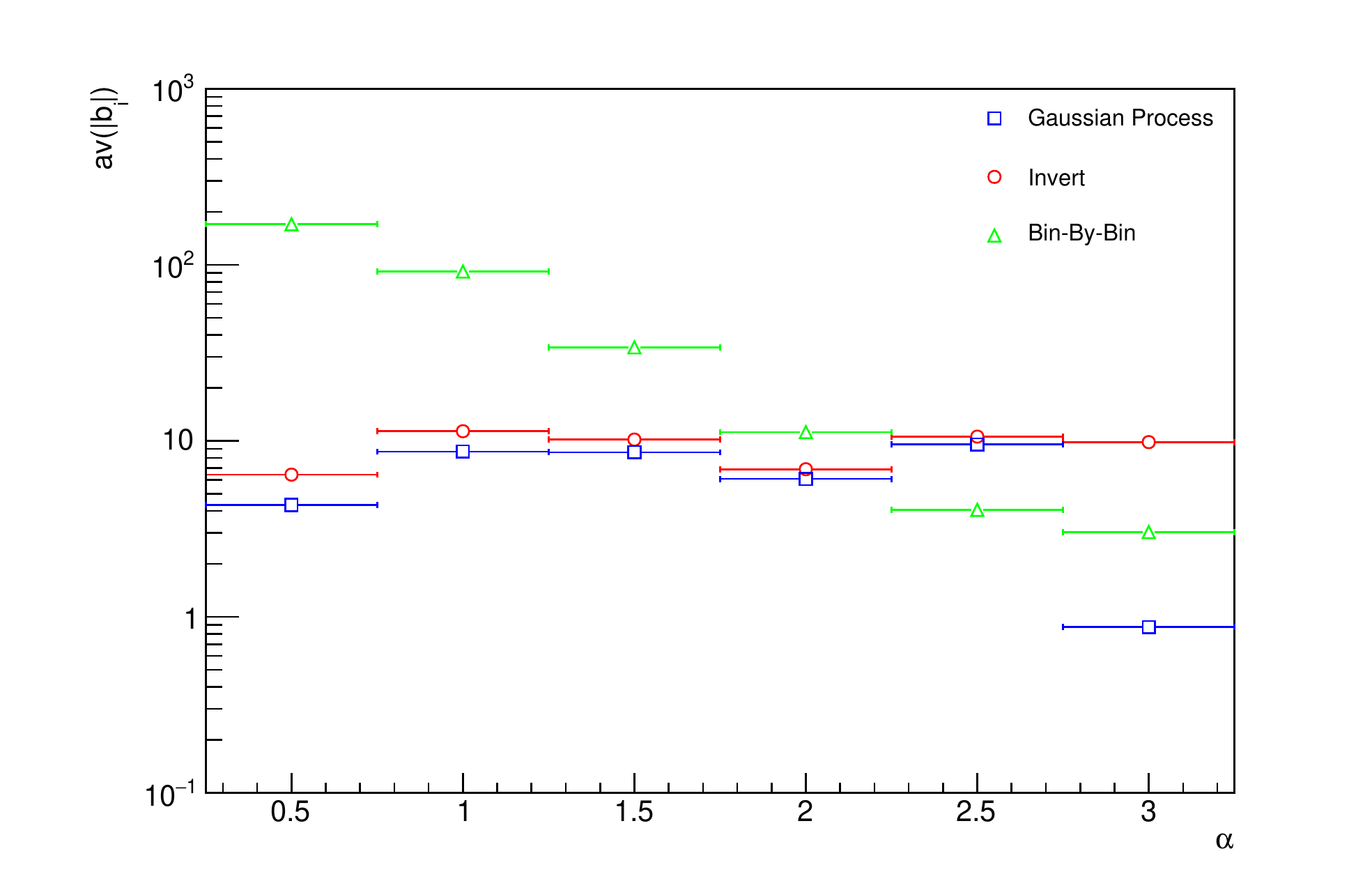} 
\caption{\label{fig:bias_bimodal_oth} 
Comparison of the estimated bin-averaged bias of non-tuneable and non-regularised methods 
when unfolding the distorted bimodal model with a response matrix obtained from the undistorted model. Results are shown for various distortion
strengths, which increase in distortion with decreasing values of $\alpha$ (See Fig.~\protect{\ref{fig:scan_bimodal}}).}
\end{center}
\end{figure}

Fig.~\ref{fig:bias_bimodal_cov} shows the performance of non-tuneable and non-regularised methods (Gaussian Process, Matrix Inversion, Bin-by-bin). The Gaussian Process and Matrix inversion methods demonstrate
to be relative robust against distortion, with a minimal dependence of the average bias on the distortion parameter $\alpha$, whereas the Bin-by-bin method shows a dependence that is comparable to that of the
tuneable regularised methods. The robustness of the Gaussian Process method is a feature of its regularisation method, which relies only on smoothing, just like TUnfold. In terms of absolute performance, 
Matrix Inversion and Gaussian Process are on par with TUnfold for medium to large distortion scenarios ($\alpha=0.5,1,1.5,2$), while Gaussian Process unfolding is the clear winner for the low distortion $\alpha=3$ scenario, 
likely due to the strongly Gaussian nature of the undistorted bimodal model.

\section{Summary and conclusions}
\label{sec:discussion}

We have used the RooFitUnfold package, which provides a common framework to evaluate and use different unfolding algorithms, to evaluate the performance
of seven unfolding algorithms used in HEP in terms of bias and coverage. The ability of RooFitUnfold to tune the regularisation strength of regularised unfolding methods
with a common set of optimisation strategies has 
facilitated 
the comparative studies presented in the paper.

For the methods with a tuneable regularisation strength we observed that an optimisation of that strength solely on the smallest MSE does 
not automatically result in good coverage. For the studied example distributions we observed that the inclusion of a bin-averaged coverage criteria in the regularisation strength optimisation resulted in uniformly good coverage for all bins for the methods with a tuneable regularisation strength. While the unfolding bias always depends on the assumed true distribution, we observed for the bimodal example that this dependence was strongest for methods that regularise using the assumed true distribution. 

\section*{Acknowledgments}
We would like to thank Tim Adye for his support. We would also like to thank the ROOT team for their help. This research is supported by the INSIGHTS Innovative Training Network under Horizon 2020 MSCA Grant 765710. This research has the support of Department of Energy Office of Science Grant DE-SC0019032. Finally, we would like to thank the Royal Holloway University of London, Tufts University, the DESY and Nikhef institutes and CERN for giving us the opportunity to pursue this endeavour.


\clearpage
\bibliography{main.bib}

\end{document}